\pdfoutput=1
\documentclass[10pt,sigconf]{acmart}
\usepackage{bbm}
\usepackage{mathrsfs}
\usepackage{graphicx}
\usepackage{amsfonts}
\usepackage{flushend}
\usepackage{amsmath}
\usepackage{graphicx}
\usepackage{subfigure}
\usepackage{epsfig}
\usepackage{latexsym}
\usepackage{amsfonts}
\usepackage{amssymb}
\usepackage{paralist}
\usepackage{xspace}
\usepackage{mathrsfs}
\usepackage{amssymb}
\usepackage{color}
\usepackage{flushend}
\usepackage{paralist}
\usepackage{multirow}
\usepackage{dsfont}
\usepackage{multicol}
\usepackage{etoolbox}
\usepackage{courier}
\usepackage{booktabs}
\usepackage{pifont}
\usepackage{setspace}
\usepackage{caption}
\usepackage{booktabs}
\usepackage{amscd}
\usepackage{relsize}
\usepackage{listings}
\usepackage{color}
\usepackage{enumitem}
\usepackage{url}
\usepackage{cancel}
\usepackage[ruled]{algorithm2e}
\usepackage{textcase}
\usepackage{dblfloatfix}
\usepackage[flushleft]{threeparttable}
\usepackage{color,soul}
\usepackage{xcolor}

\def\ie{i.e.,\xspace}

\def\etc{etc}
\def\eg{e.g.,\xspace}

\definecolor{dkgreen}{rgb}{0,0.6,0}
\definecolor{gray}{rgb}{0.5,0.5,0.5}
\definecolor{mauve}{rgb}{0.58,0,0.82}

\newcommand{\oursystem}{\texttt{UPS+}\xspace}
\newcommand{\ubeacon}{uBeacon\xspace}
\newcommand{\ubeacons}{uBeacons\xspace}
\newcommand{\cbeacon}{cBeacon\xspace}
\newcommand{\cbeacons}{cBeacons\xspace}

\DeclareMathOperator*{\argmax}{argmax\,}

\graphicspath{{figs/}}

\renewcommand\footnotetextcopyrightpermission[1]{} % removes footnote with conference information in first column
\usepackage{etoolbox}% http://ctan.org/pkg/etoolbox
\makeatletter
\patchcmd{\maketitle}{\@copyrightspace}{}{}{}
\makeatother

\begin{document}

\title{Rebooting Ultrasonic Positioning Systems for Ultrasound-incapable Smart Devices}
\author{Qiongzheng Lin$^\dagger$, Zhenlin An$^\dagger$, Lei Yang}
\affiliation{%
  \institution{$^\dagger$Co-primary Authors} 
  \institution{Department of Computing}  
  \institution{The Hong Kong Polytechnic University}  
}
\email{{lin, an, young}@tagsys.org}

\begin{abstract}

An ultrasonic Positioning System (UPS) has outperformed RF-based systems in terms of its  accuracy for years. However, few of the developed solutions have been deployed in practice to satisfy  the localization demand of today's smart devices, which  lack ultrasonic sensors and were considered as being ``deaf'' to ultrasound.
A recent finding demonstrates that ultrasound may be audible to the smart devices under certain conditions due to their microphone's nonlinearity. Inspired by this insight, this work revisits the ultrasonic positioning technique and builds a practical UPS, called \oursystem, for ultrasound-incapable smart devices. The core concept is to deploy two types of indoor beacon devices, which will advertise ultrasonic beacons at two different ultrasonic frequencies respectively.  Their superimposed beacons are shifted to a low-frequency by virtue of the nonlinearity effect at the receiver's microphone. This underlying property functions as an implicit ultrasonic downconverter without throwing harm to the hearing system of humans.  We demonstrate \oursystem, a fully functional UPS prototype, with centimeter-level localization accuracy using custom-made beacon hardware and well-designed algorithms. 
\end{abstract}




\maketitle

\section{Introduction}
\label{section:introduction}

Tracking smart devices, such as phones or wearables, inside buildings where the GPS is not available has become a growing business interest. Indoor localization enables users to navigate indoor spaces similar to the function provided by GPS for outdoor environments.  It prompts a series of key mobile applications, namely, indoor navigation (\eg malls, factories, and airports), augmented reality, location-aware pervasive computing, advertising, and social networking.

\begin{figure}[t!]
  \centering
  \includegraphics[width=0.8\linewidth]{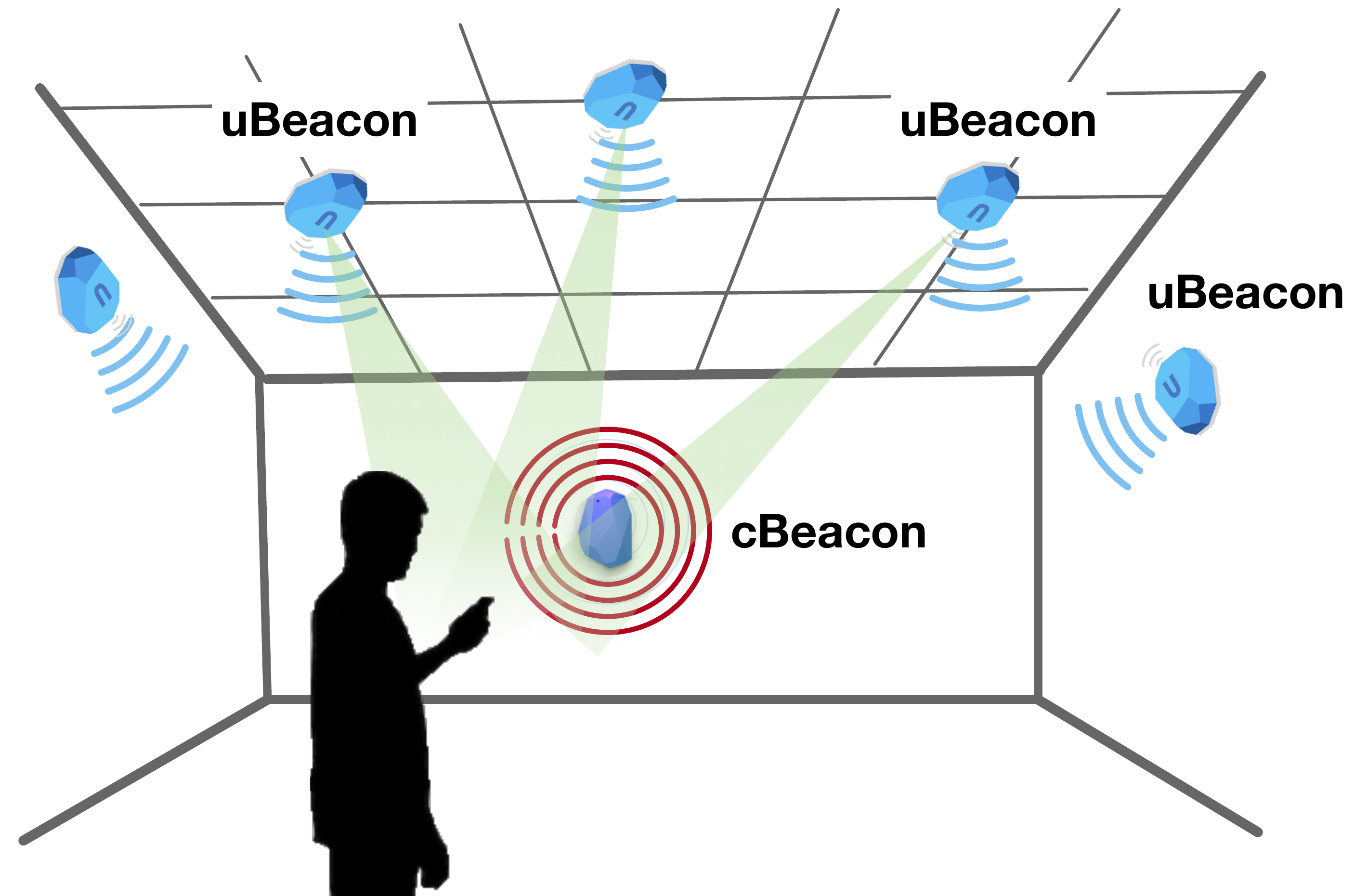}
    \vspace{-0.2cm}
   \caption{System architecture. \textnormal{Multiple \ubeacons are deployed as location anchors for the trilateration, whereas a single \cbeacon is installed for ultrasonic downconversion.}}
  \label{fig:achitecture}
    \vspace{-0.3cm}
\end{figure}

Many efforts have been exerted  to deliver an accuracy of tens of submeters for indoor localization. However, only a few of these solutions have actually reached today's mobile devices because previous works have failed in one of two categories at a high level; that is, they either require mobile devices to be equipped with specialized sensors, such as accelerators~\cite{shu2015last}, magnetic sensors~\cite{haverinen2009global}, LEDs~\cite{zhang2017pulsar,zhu2017enabling}, RFIDs~\cite{wang2007rfid,wang2013dude,wang2013rf,ma20163d,ma2017minding}, and WiFi~\cite{xiong2013arraytrack}, \etc; or, they require exhaustive fingerprinting of the environment to learn the spatial distribution of signal characteristics, such as FM ~\cite{chen2012fm}, WiFi~\cite{bahl2000radar,chintalapudi2010indoor}, Bluetooth~\cite{johnson2012localization}, sound or light~\cite{azizyan2009surroundsense,tan2013sound}, and Zigbee~\cite{gao2013zifind}, \etc. Ultra-wide band (UWB) (\eg WiTrack~\cite{adib20143d}) can achieve cm- (even mm-)  level accuracy but requires GHz bandwidth, which is not allowed in practice due to the spectrum regulations. On the contrary, as the rapid development of smart technology, The demand for highly accurate indoor localization services has become a key prerequisite in some markets, thereby resulting in a growing business interest.  For example, finding the AirPod in a head, tracking user's arms, writing in the air, pinpointing items in VR/AR systems, and so on.

\begin{figure*}[t!]
  \centering
  \includegraphics[width=0.7\linewidth]{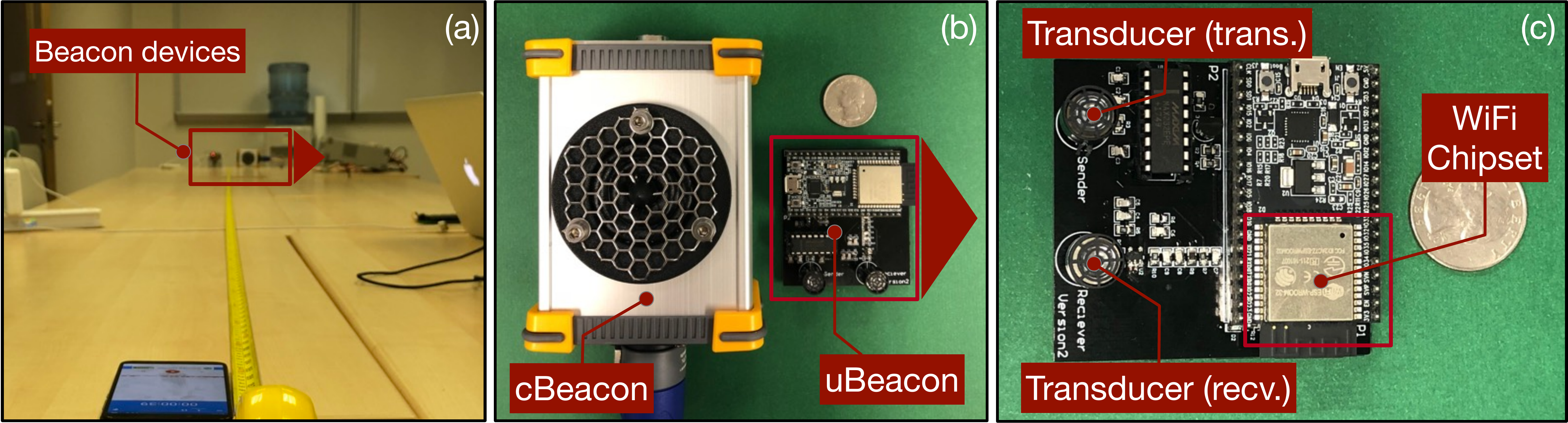}
   \caption{\oursystem experimental platform. \textnormal{(a) Experimental scenario; (b) two beacon devices, \cbeacon and \ubeacon; and (c) zoom-in image of the circuit board of the \ubeacon.}}
  \label{fig:devices}
\end{figure*}

In this work, we revisit a classical indoor localization solution, namely, the \emph{ultrasonic positioning system} (UPS), which utilizes the ultrasonic sound as the ranging media.  UPS  outperforms RF-based systems in terms of accuracy.  For example, MIT launched  Project Oxygen~\cite{oxygen} in 2004. In this project, a pioneering UPS named  Cricket ~\cite{priyantha2000cricket} was  developed to provide centimeter-level accuracy location service. Subsequently, a large number of follow-up works~\cite{want1992active,priyantha2001cricket,smith2004tracking,lazik2012indoor,lazik2015ultrasonic,borriello2005walrus,tarzia2011indoor,peng2007beepbeep,hazas2002novel,mccarthy2006accessible,yang2011detecting,mandal2005beep} have further utilized sound or ultrasound for  ranging or localization (see $\S$\ref{section:related-work}).  However, despite the extensive efforts exerted on UPS research, only a few of the resulting solutions have been actually deployed in practice at present. For example,  Cricket project ceased updating 10 years ago. One of the important reasons why UPSs do not receive enough attention today is that they suffer from a serious defect; that is, today's smart devices lack ultrasonic sensors, and therefore, cannot receive ultrasonic beacons from UPSs~\cite{yang2011detecting,lazik2015ultrasonic,peng2007beepbeep}. To bridge this gap, many follow-up works~\cite{yang2011detecting,lazik2015ultrasonic,peng2007beepbeep} have advised  transmitting beacons at the spectrum of $20\sim 22$ kHz, where $22$ kHz is the upper bound of currently available microphones. The sound within this band is supposedly inaudible to the majority of humans.  Nevertheless, it is still harmful to infants and pets who are hyper-sensitive to the high-frequency sounds. Moreover, the signals near the upper bound  may be seriously distorted and attenuated due to the non-ideal transition band of  the low-pass filter in the microphone.

We present \oursystem, an enhanced UPS that can provide sub-centimeter indoor localization to current smart devices and operate absolutely at ultrasonic spectrum, which is considerably beyond the hearing system of humans and pets. Our key innovation is the \emph{renewed efforts} on promoting traditional UPSs to  become serviceable to ultrasound-incapable receivers.   A recent finding shows that a combination of two ultrasounds at two frequencies (\eg $f_1$ and $f_2$) may get shifted to a lower differential frequency (\eg $|f_1-f_2|$)  when they arrive simultaneously at a microphone~\cite{zhang2017dolphinattack,roy2017backdoor,roy2018inaudible}. 
Toward this hardware property as a natural downconversion approach,  we ``pull down'' ultrasonic beacons to the audible spectrum, which the receiver can process.  To this end,  \oursystem adopts a \emph{heterogeneous architecture} that consists of two types of custom-made beacon devices: \ubeacon and \cbeacon, as shown in Fig.~\ref{fig:achitecture}. They broadcast beacons at two \emph{ultrasonic frequencies},  thereby ensuring that the beacon signals will not disturb humans or pets. In particular, a large number of  battery-supplied \ubeacons are deployed as location anchors, whereas  a single cable-powered \cbeacon is installed for the ultrasonic downconversion. Finally, periodic beacons from \ubeacons are downconverted by the beacon signals from the \cbeacon.  After capturing the beacons from at least four \ubeacons, the receiver computes the time of arrival and then locates itself via trilateration.

The nonlinearity effect has been verified and demonstrated in various acoustic attacks~\cite{roy2017backdoor,zhang2017dolphinattack,roy2018inaudible}. However,  whether such property can be used for indoor localization remains unclear, unless the following three concerns are addressed:

$\bullet$  \emph{How can we deal with the frequency-selectivity?} Several copied ultrasounds with the same frequency but propagated along different  paths may induce constructive or destructive interference at the receiver, thereby leading to the frequency selectivity issue.  The use of broadband acoustic signals (\ie chirps) is considered as an effective means to address this challenge~\cite{lee2015chirp,lin2017tagscreen,hazas2002novel,lazik2012indoor}. However,  a \ubeacon is made of an ultrasonic transducer and only has a bandwidth of $2$~kHz, which is barely  adequate to deal with the selectivity. In this work, we propose the technique of \emph{dynamical chirp spread spectrum} (DCCS) technique, which spreads the  single-tone pulses of the \ubeacons over the \cbeacon's broadband chirps.

$\bullet$ \emph{How can the beacons advertised from multiple beacon devices be distinguished?} Trilateration requires the receiver to acquire beacons from at least four \ubeacons; thus,  multiple beacon access is an unavoidable issue. Frequency division multiple address (FMDA) is typically adopted in previous UPSs. However, it fails in our scenario because each \ubeacon sweeps an unpredictable dynamic band that depends on the receiver's location, which causes
 the downconverted beacons to overlap in the frequency-domain at the receivers.  Instead, \oursystem is driven by a two-level multiple access mechanism. The receiver initially locates itself in a large space through the chirp slope of the \cbeacon and then decodes the \ubeacon's ID from its scheduled beacons.

$\bullet$  \emph{How can the receiving energy of downconverted beacons be enhanced?} The nonlinearity effect is observed at the second-order harmonics, which has an amplitude that is less than the fundamental signals. Consequently, the signal-to-noise ratio (SNR) of the beacon at the receiver side in \oursystem is less than that in a traditional UPS at the same position.  We alleviate this problem through the following efforts. On the transmitting side, a complicated custom-made driver circuit is designed to double the output power of the transducer. On the receiving side, a noise reduction algorithm is specially redesigned to turbocharge weak beacons by taking advantage of the dual-microphones in smart devices.

\textbf{Summary of Results.} Fig.~\ref{fig:devices} shows the experiment platform and the prototype.   Our evaluation is performed on $8$ types of off-the-shelf smart devices (\ie five smartphones, an iPad, an iWatch and a pair of AirPods). The results demonstrate that \oursystem can fully utilize the $22$ kHz bandwidth for chirps spread on these devices. It performs localization with median and $90^{th}$ percentile errors of $4.59$ cm and $14.57$ cm, respectively. Such accuracy matches or even exceeds some previous UPSs.  The effective range (median error $\leq 20$ cm) of \ubeacon equipped with three-transducers is around $6$ m as equivalent to that of normal UPSs.

\textbf{Contribution}. This work presents \oursystem, the first system that operates at an ultrasonic spectrum but remains serviceable in currently available ultrasound-incapable smart devices. The design of \oursystem introduces three key innovations. First, it presents a centralized device called \cbeacon to boost  previous UPSs. Second, it uses the nonlinearity effect of microphone systems as an ultrasonic downconverter. Third, it spreads single-tone pulses with dynamic chirps in the air and turbocharges downconverted beacons at the receiving side.  The study also presents a prototype implementation and evaluation of \oursystem, thereby demonstrating its accuracy in localizing smart devices in our office.

\section{Related Work}
\label{section:related-work}
This work touches upon many topics related to indoor localization, nonlinearity effect and noise reduction. These areas already have large bodies of research; thus, our review  focuses primarily on closely related works.

\textbf{(a) Sound-based Indoor Localization}:
Sound-based indoor localization can be classified under broad categories of range-based~\cite{priyantha2000cricket,harter2002anatomy,hazas2002novel,hazas2003high,mccarthy2006accessible,mandal2005beep,lazik2012indoor,lazik2015ultrasonic,filonenko2010investigating,peng2007beepbeep,nandakumar2015contactless,lazik2015alps} and range-free~\cite{rossi2013roomsense,lorincz2005motetrack,tarzia2011indoor,lorincz2005motetrack,lu2009soundsense,azizyan2009surroundsense,tan2013sound,tung2015echotag} methods. Here, we focus only on range-based methods.  We refer to \cite{panchpor2018survey} for a comprehensive survey on various indoor localization techniques.  This category of solutions computes distances based on how long sound takes to propagate between a sender and a receiver~\cite{priyantha2000cricket,harter2002anatomy,hazas2002novel,hazas2003high,mccarthy2006accessible,mandal2005beep,lazik2012indoor,lazik2015ultrasonic,filonenko2010investigating,peng2007beepbeep,nandakumar2015contactless}.  For example, Active Bats~\cite{harter2002anatomy} and Cricket~\cite{priyantha2000cricket} are pioneering range-based localization systems that use ultrasonic beacons. Dolphin~\cite{hazas2002novel} presents a new design for ultrasonic transmitters and receivers.   BeepBeep~\cite{peng2007beepbeep} uses the same range approach to estimate the distance between two cellular phones. \cite{yang2011detecting} attempts to identify the location of a mobile phone in a car using vehicle-mounted audio speakers. ApneaApp~\cite{nandakumar2015contactless} uses an estimated range  to track human breath.  Pulse Compression(PC)~\cite{lazik2012indoor} is the closest to our work but it uses $19\sim 20$ kHz band, which may still be sensitive for infants or pets. By contrary, our beacons are advertised at $50$ kHz or higher, which results in zero noise pollution to indoor creatures.  Moreover, our work uses dynamic chirps ($\S$\ref{section:toa}) to spread the spectrum instead of the static chirps used in~\cite{lazik2012indoor}.  In summary, previous works either  require ultrasonic receivers or advertise audible beacons. Additional comparisons are provided in $\S$\ref{section:evaluation}.

\textbf{(b) Nonlinearity Effect:} Nonlinearity has been explored for many purposes~\cite{roy2017backdoor,zhang2017dolphinattack,roy2018inaudible} in recent years.  Backdoor~\cite{roy2017backdoor} constructs an acoustic (but inaudible) communication channel between two speakers and a microphone over ultrasound bands with the nonlinearity effect. DolphinAttack~\cite{zhang2017dolphinattack} and LipRead~\cite{roy2018inaudible} utilize the nonlinearity effect to send inaudible commands to voice-enabled devices such as Amazon Echo.  Our system is inspired by these previous works. However, we use this effect for a different purpose and face challenges that vary from those in previous studies.

\textbf{(c) Noise Reduction:} Finally, several works~\cite{jeub2012noise,chen2017background,martin2001noise,jeub2011robust} from the speech field have motivated us to design a dual-microphone enabled turbocharging algorithm. In contrast to previous works, we use the features of beacon signals to determine their absence.

\section{Towards Nonlinearity as an ultrasonic downconverter}

In this section, we provide the background of nonlinearity effect and describe how \oursystem can take advantage of this phenomenon for ultrasonic downconversion.

\subsection{Primer on Microphone System}

\begin{figure*}[t]
  \centering
  \subfigure[Pulse+CW]{
    \includegraphics[width=0.33\linewidth]{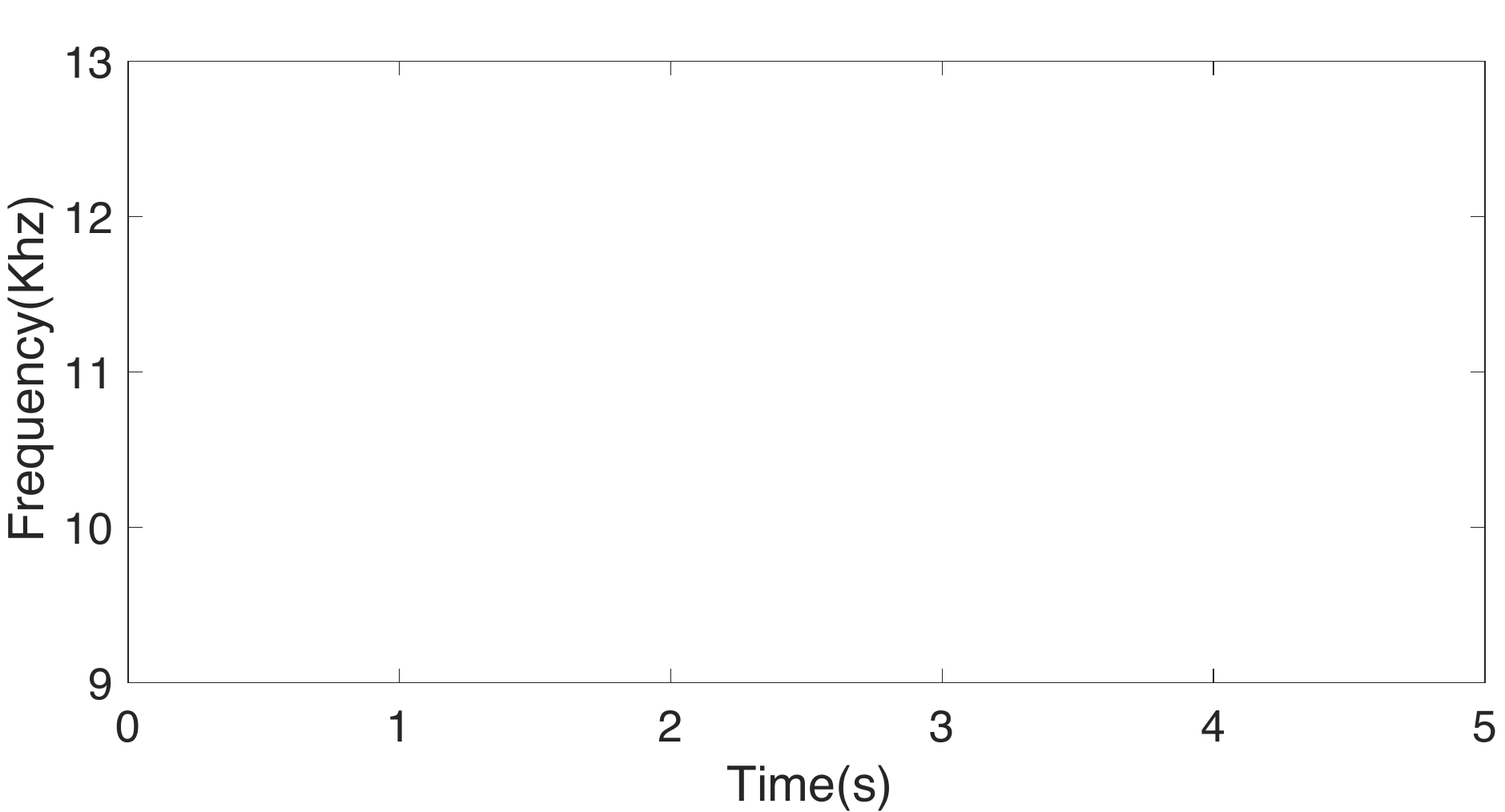}
    \label{fig:downconverted-beacon-1} 	
  }%
  \subfigure[CW+Chirps]{
    \includegraphics[width=0.33\linewidth]{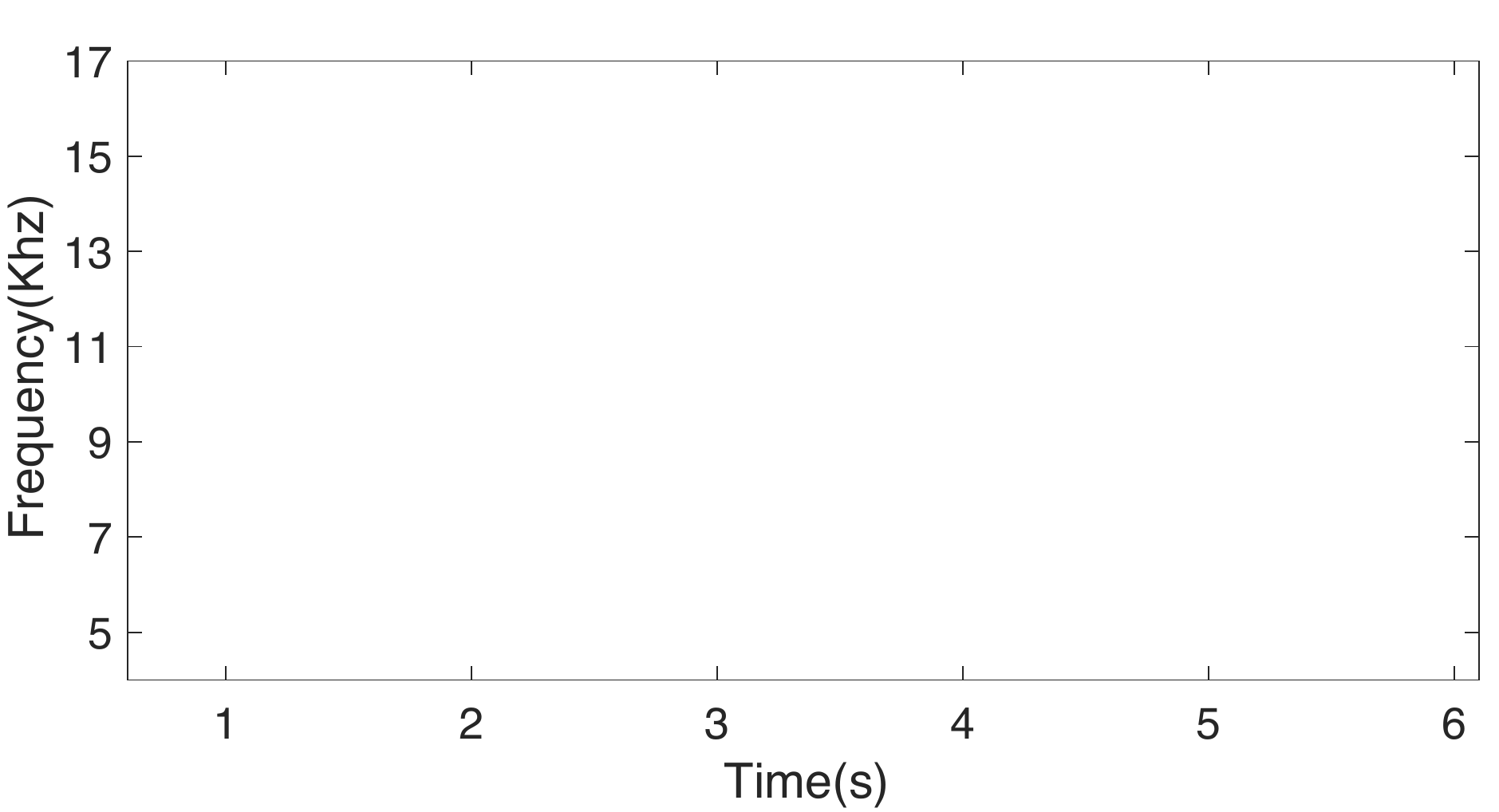}
     \label{fig:downconverted-beacon-2}
  }%
   \subfigure[Pulse+Chirps]{
    \includegraphics[width=0.33\linewidth]{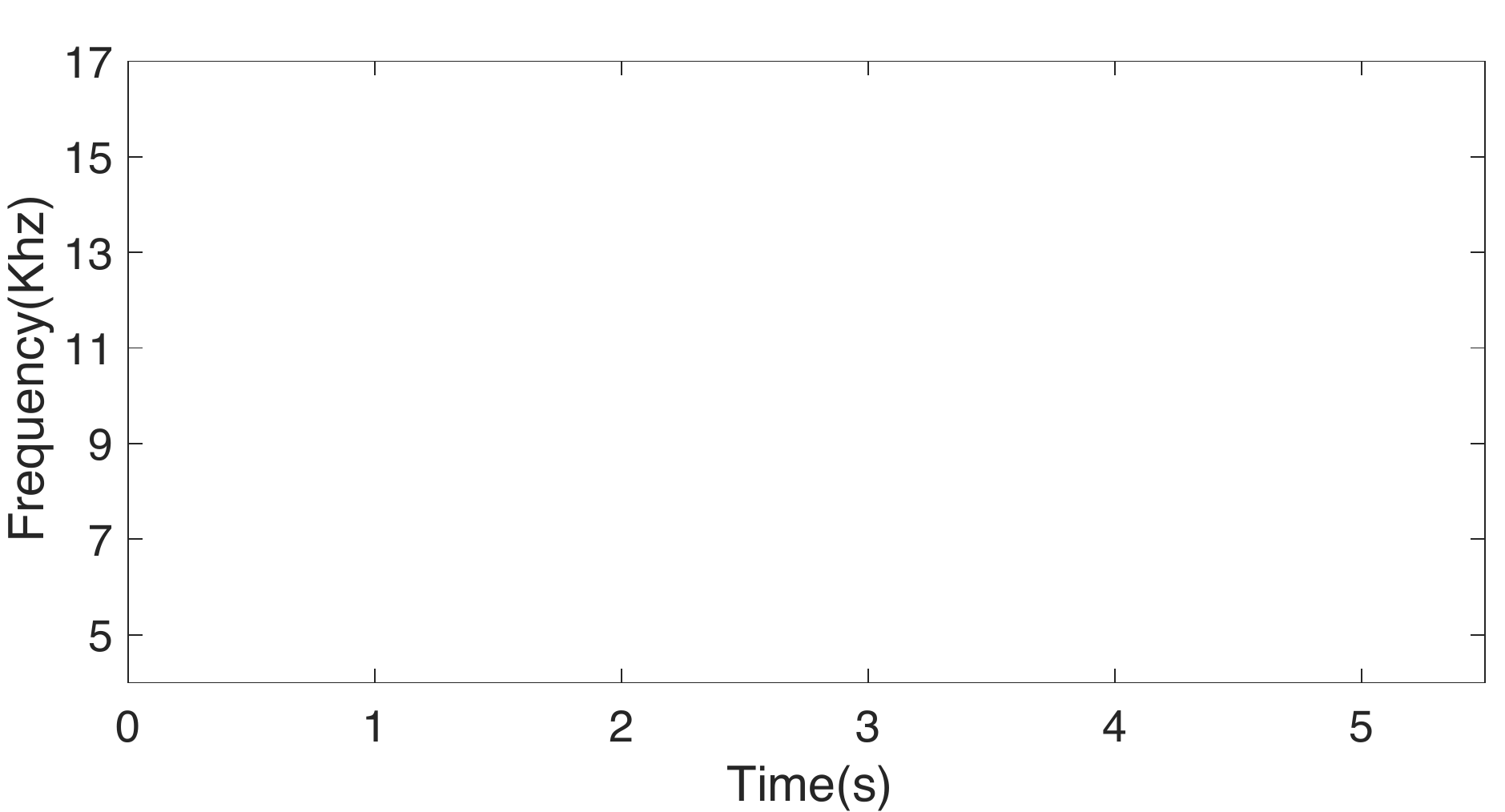}
     \label{fig:downconverted-beacon-3}
  }
  \vspace{-0.1cm}
  \caption{Downconverted beacons in the frequency-domain. \textnormal{The built-in recorder of an iPhone 8 is used to capture ultrasonic beacons with different settings: (a) the \ubeacon advertises pulses at $40$ kHz whereas the \cbeacon transmits a single-tone continuous wave (CW) at $50$ kHz; (b) the \ubeacon advertises a CW at $40$ kHz while the \cbeacon transmits continuous chirps from $45\sim 55$ kHz; and (c) the \ubeacon advertises pulses at $40$ kHz, whereas the \cbeacon transmits chirps from $45\sim 55$ kHz.}}
  \vspace{-0.3cm}
\end{figure*}

Today's smart devices adopt a generic pipeline for sound processing. After being captured by the microphone, sound is first magnified by \emph{amplifiers}. To capture audible sounds, the microphone system is designed to be sensitive to the spectrum of $0\sim 22$ kHz by using a \emph{low-pass filter} (LPF) to remove sounds that are higher than $22$ kHz, even when they are recorded by the microphone. Finally, the sampling rate of the analog-to-digital converter is typically $44.1$ kHz, and the digitized signal's frequency is limited to below $22$ kHz according to the Nyquist sampling theorem. All modules within the microphone system are supposed to be linear; thus, the output signals are linear combinations of the input. However,  the sound amplifiers exhibit strong \emph{nonlinearity} outside the operating band.	 In particular, if $f>25$ kHz, then the net recorded sound $S_\text{out}$ can be expressed in terms of the input sound $S$ as follows:
\begin{equation}\small
	S_{\text{out}}(t) \vert_{f>25}= \sum_{i=1}^{\infty} G_i S(t) = G_1S(t) + G_2 S^2(t) + G_3 S^{3}(t) + \cdots 
\end{equation}
where $G_i$ is the gain for the $i^{th}$ item. The amplifier creates new frequencies (\ie harmonics) due to the nonlinearity property~\cite{horowitz1989art}.

\subsection{Nonlinearity Effect}
To operate the microphone system in an ultrasonic band, we use two off-the-shelf ultrasound speakers to simultaneously play two sounds $S_1(t)=A_1\cos(2\pi f_1 t)$ and $S_2(t)=A_2\cos(2\pi f_2 t)$ at frequencies of $f_1$ and $f_2$ respectively, where $f_1$ and $f_2>22$ kHz.  After magnification, the output $S_\text{out}$ can be modeled as follows:
\begin{equation}\small\label{eqn:sound-equation}
\begin{split}
	S(t) =& G_1(S_1+S_2) + G_2(S_1+S_2)^2 + G_2(S_1+S_2)^3+\cdots
\end{split}
\end{equation}
We discuss only the second-order term by manipulating the input signal $S$ because the third- and higher-order terms have an extremely weak gain and can be disregarded.  We expand the second-order item as follows:
\begin{equation}\small
\begin{split}
	 &G_2(S_1+S_2)^2= \frac{1}{2}(G_2A_1^2+G_2A_2^2)-\frac{1}{2}(\cos(2\pi f_1 t)+\cos(2\pi f_2 t))\\
	 &+G_2A_1A_2\cos(2\pi (f_1+f_2) t)- G_2A_1A_2\cos(2\pi (f_1-f_2) t)
\end{split}\raisetag{\baselineskip}
\end{equation}
Interestingly,  four new frequencies (\ie $2f_1$, $2f_2$, $f_1-f_2$ and $f_1+f_2$) are created after magnification. All the ultrasonic frequencies at $f_1$, $f_2$,  $2f_1$, $2f_2$ and $f_1+f_2$ are filtered out due to the LPF's cut off at $22$ kHz. However, $(f_1-f_2)$ remains.  Translating to actual numbers, when $f_1=50$ kHz and $f_2=40$ kHz, the microphone will output acoustic signals at $10$ kHz. The net effect is that a completely inaudible frequency is recorded by unmodified off-the-shelf microphones, thereby offering \emph{natural downconversion} for ultrasonic signals. For brevity, we extract the downconverted sound signal denoted by $S_{\downarrow}(t)$  as our target  signals: 
\begin{equation}\label{eqn:down-beacon}\small
	S_{\downarrow}(t) = G_2A_1A_2\cos(2\pi (f_1-f_2) t) = A_{\downarrow}\cos(2\pi (f_1-f_2) t) 
\end{equation}
where $A_{\downarrow}=G_2A_1A_2$.  The feasibility and accuracy of the nonlinearity effect on existing microphone systems have been completely verified and demonstrated in recent previous works. Here, we omit to conduct our own verification  and refer readers to the  results in~\cite{zhang2017dolphinattack,roy2017backdoor,roy2018inaudible}. Especially, \cite{zhang2017dolphinattack} tests dozens of devices to show that the nonlinearity is a ubiquitous effect across the currently available smart device.

\subsection{Ultrasonic Downconversion}
Nonlinearity effect has been  considered a type of ``pollution'' or a security ``back door'' in previous works.  Instead, we explore this underlying physical property as a novel and positive \emph{downconverter} to engineer a practical UPS.  Such downconverter does not require receivers to be equipped with any ultrasonic sensors. It offers two clear advantages. First, downconversion is generated by the hardware property of electronic amplifiers, and thus, it will not affect humans or animals. Second, the nonlinearity effect can downconvert ultrasonic beacons into any frequency below $22$ kHz by manipulating two ultrasonic speakers, thereby implicitly providing up to $22$ kHz broadband. Subsequently, we will explore this effect to design \oursystem.

\section{SYSTEM DESIGN}

We call the ultrasonic signal \emph{beacon} and the devices that can transmit ultrasonic beacons, \emph{beacon device}. Our fundamental concept is to deploy many  beacon devices in the target space and utilize the nonlinearity effect to downconvert their beacons into an audible spectrum that can be recognized by smart devices.  We present the system design at a high-level in this section.

\subsection{Dilemma in Engineering UPS+} 

To take advantage of the nonlinearity, the straightforward design should equip each beacon device with two ultrasonic speakers, which transmit ultrasound at two different frequencies. However, this design is \emph{uneconomical} and \emph{impractical} in engineering practices. \emph{Our engineering  philosophy is to utilize commercial off-the-shelf components to build the entire system and to benefit from the economies of scale by being cost-effective.} In the present market, two types of commercial ultrasonic components, namely, \emph{transducers} and \emph{speakers}, are available as listed in Table.~\ref{tab:ultrasonic-devices}.

\begin{table}[!t]
  \centering\small
  \caption{Comparisons of ultrasonic components}
  \label{tab:ultrasonic-devices}
  \vspace{-0.3cm}
  \begin{tabular}{rccccc}
    \toprule
     \textbf{Type} &\textbf{Size($cm^2$)} &\textbf{Price} & \textbf{BW} & \textbf{Rng.} &\textbf{Supply}  \\
    \hline
     Transducer & $1.6\times 1.6$&$1$ \$&$2$ kHz &  $8$ m & Battery\\
	 Speaker & $9\times 12$ & $50$ \$ & $200$ kHz &  $50$ m & Cable\\
    \bottomrule
  \end{tabular}
  \vspace{-0.3cm}
\end{table}

\begin{itemize}[leftmargin=*]	
	\setlength{\parskip}{0pt}
  	\setlength{\itemsep}{0pt plus 1pt}
	\item \textbf{Transducers}: Ultrasonic transducers (aka ultrasonic sensors) can convert AC into ultrasound, or vice versa. They are small, low cost ($1$ \$), and power-saving, but suffer from a limited range (\ie max at $8$ m) and a narrow band ($\sim 2$ kHz). The majority of previous UPSs (\eg Cricket) use transducers to build beacon devices. 
	\item \textbf{Speaker}: Ultrasonic speakers are bulky and expensive ($50$ \$), but have a broadband of $200$ kHz (\ie $0\sim 200$ kHz) and a propagation range of up to $50$ m. These speakers were used as anchors in PC~\cite{lazik2012indoor}.
\end{itemize}
A comparison of their characteristics presents an engineering dilemma in component selection; that is, deploying a large number of speakers is extremely expensive, whereas cheap transducers fail to satisfy our broadband demand. 

\subsection{System Architecture}

To overcome the dilemma,  we design a \emph{heterogeneous architecture} that  jointly uses the two ultrasonic components by inventing two types of beacon devices.
\begin{itemize}[leftmargin=*]\setlength{\parskip}{0pt}\setlength{\itemsep}{0pt plus 1pt}
	\item \textbf{\ubeacon}: A \ubeacon is composed of an ultrasonic transducer, operating at a single tone (\eg  $40$ kHz) with a directivity of $30^\circ$.   The total cost of a \ubeacon is about $5$\$. We can deploy \ubeacons at a large scale due to its low cost. \ubeacons act as location anchors, whose locations are known ahead of time to receivers.
	\item \textbf{\cbeacon}: A \cbeacon is composed of a powerful ultrasonic speaker. A single \cbeacon is sufficient to cover a large target space (\eg a lecture room) and shared by many \ubeacons since the speaker is wideband and long-range.  Powered by electric cables, a \cbeacon produces \emph{continuous} ultrasonic signals and functions for downconversion. It can be attached anywhere because its location is irrelevant to the localization results.
\end{itemize}

Such heterogeneous architecture achieves a trade-off between the cost and effectiveness.  In particular, we can immediately  stop the potential interference to other recording activities (\eg recording voice using smart phones.)  by turning off  the single \cbeacon, rather than stopping distributed \ubeacons one by one; or awaken \cbeacon for localization when needed.

\subsection{Timing for Downconversion}
\begin{figure}[t!]
  \centering
  \includegraphics[width=0.9\linewidth]{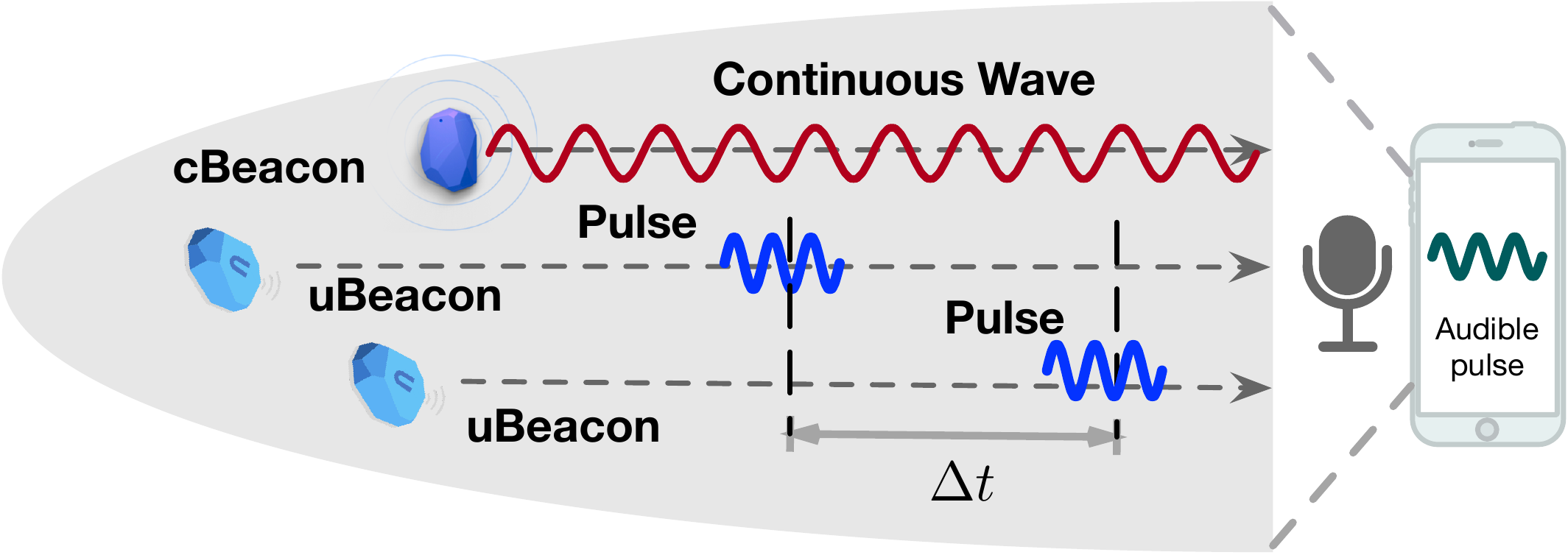}
   \caption{Downconversion timing}
  \label{fig:beacon-design}
    \vspace{-0.3cm}
\end{figure}

We assign two different ultrasonic frequencies to the \ubeacon and the \cbeacon, and they are deployed in different locations. The receiver can detect a downconverted beacon \emph{only} when their ultrasonic beacons arrive simultaneously at the receiver. A natural question is how \oursystem synchronizes the arrivals of two types of beacons. To do so, the \cbeacon persistently transmits a \emph{continuous wave} (CW), whereas the \ubeacons advertise short beacons every a few milliseconds. Fig.~\ref{fig:beacon-design} uses a toy example to explain the design. Given that signals from the \cbeacon arrive continuously, downconversion occurs only when  a \ubeacon's signal arrives at the receiver.  In this manner, timing is dependent only on the signal arrival of the \ubeacon and irrelevant of the \cbeacon. Furthermore,  we demonstrate the timing through an actual experiment in Fig.~\ref{fig:downconverted-beacon-1}, where the \ubeacon advertises pulses at $40$ kHz, whereas the \cbeacon transmits a single-tone CW at $50$ kHz. We observe an apparent pattern of pulses at $10$ Hz every second similar to the \ubeacon's advertising pattern at $40$ kHz, although the \cbeacon never stops its signals. 

\subsection{Trilateration with ToA in UPS+}

Suppose that the $i^{th}$ \ubeacon is located at position $U_i$ and transmits a beacon at time $t_s$. The receiver  detects the arrival of this beacon at time $t_i$. Thus, the beacon takes $t_i-t_s$  seconds to propagate in air. We can compute the \emph{pseudo-range} as follows:
\begin{equation}\label{eqn:pseudo-range}
	 c\times (t_i-t_s) = c\times t_b + |U_i-P|
\end{equation}
where $c$ is the speed of sound, $P(x,y,z)$ is the receiver's location, and $t_b$ is the clock difference between the receiver and the \ubeacons. All the \ubeacons are assumed to be well synchronized in time and advertise beacons at time $t_s$.  $c$, $t_i$ and $U_i$ are known to the receiver; hence, the preceding equation contains four unknowns, including three coordinate variables in $P(x,y,z)$ and $(t_b+t_s)$. The receiver's OS kernel may introduce additional internal delays due to multi-threading. These delays can be counted to $t_b$.   To solve $P(x,y,z)$, the receiver must receive beacons from at least $4$ different \ubeacons.  This approach is  called \emph{trilateration}.  At a high level, \oursystem's trilateration has the following components.
\begin{itemize}[leftmargin=*]
	\setlength{\parskip}{0pt}
  	\setlength{\itemsep}{0pt plus 0pt}
	\item \textbf{Estimation of ToA}: The first component estimates the ToA of each beacon from recorded audio data in a complex environment (see $\S$\ref{section:toa}).
	\item \textbf{Multiple Beacon Access}: The second component identifies the source device of each beacon when multiple \ubeacons are present (see $\S$\ref{section:mba}).
	\item  \textbf{Enhancement of Beacons.} The third component enhances the SNRs of downconverted beacons at both transmitter and receiver sides (see $\S$\ref{section:enhance}).
\end{itemize}

\noindent The subsequent sections elaborate these components.

\section{Estimation of ToA}
\label{section:toa}

The key to trilateration is the estimation of ToA when the beacon arrives at the microphone. A naive solution is to allow \ubeacons to transmit single-tone pulses. Then, the receiver can estimate ToA by detecting the existence of a signal at the downconverted frequency, as shown in Fig.~\ref{fig:downconverted-beacon-1}.  However, such an approach is typically challenged by background noise, echoes, and the notorious frequency selectivity.  The spread spectrum technique is widely recognized as a good solution to conquer these challenges. In this section, we present the unique spread spectrum technique adopted in \oursystem and then introduce the  estimation approach.

\subsection{Spreading Beacons with  Chirps}

A chirp is a sinusoidal signal with a frequency that increases or decreases over time. The chirp spread spectrum (CSS)  is a spread spectrum technique that uses wideband linear frequency chirps to modulate information. CSS has been proven to be resistant against noise and multipath fading in acoustic channels~\cite{lee2015chirp,lazik2012indoor}.  Suppose the sampling rate is $f_s$ and $t_0$ is the time that the first sample is obtained. Then a periodic chirp is defined as
\begin{equation}\label{eqn:chirp}\small
	S[k]= \cos\left(2\pi (f_0+\frac{1}{2}(k\bmod K)\Delta f )t_k\right)
\end{equation}
where $t_k = t_0+(k/f_s)$ and $f_0$ is the start frequency at time $t_0$. The transmitter periodically sweeps the spectrum of $[f_0,f_0+K\Delta f]$ . The swept spectrum is divided by $K$ equal intervals in the unit of samples. On the receiver end, ToA can be computed by correlating  the received signal with a predefined  chirp template. If the correlation spikes in the middle of the reception, then it indicates a match. The position of the spike corresponds to the beginning of the chirp (\ie ToA).

\textbf{Spread Beacon in the Air.} The straightforward approach is to allow  \ubeacons to directly yield chirps.  However, a chirp signal typically sweeps a wide band (\eg $5$ kHz),  which considerably exceeds a \ubeacon's bandwidth (\ie $2$ kHz).  In \oursystem, we allow the band-wider \cbeacon to transmit chirps. Let $S_u$ and $S_c$ denote the beacons transmitted by a \ubeacon and the \cbeacon, respectively.
\begin{equation}\small
	\begin{cases}
		S_u[k] = A_u \cos(2\pi  f_u t_k)\\
		S_c[k] = A_c \cos(2\pi (f_c+\frac{1}{2}(k\bmod K)\Delta f)t_k )
	\end{cases}
\end{equation}
where the \ubeacon advertises at $f_u$ and the \cbeacon periodically sweeps the spectrum within $[f_c,f_c+K\Delta f]$.  When the above two equations are substituted into Eqn.~\ref{eqn:down-beacon}, we obtain the downconverted beacon as follows:
\begin{equation}\small \label{eqn:beacon}
		S_{\downarrow}[t_k] = A_{\downarrow}\cos\left(2\pi (f_c-f_u+\frac{1}{2}(k\bmod K)\Delta f) t_k\right)
\end{equation}
A comparison between Eqn.~\ref{eqn:beacon} and Eqn.~\ref{eqn:chirp} clearly indicates that the downconverted beacon is in the form of a chirp, but the sweeping shifts to the downconverted spectrum, \ie  $[(f_c-f_u), (f_c-f_u)+K\Delta f]$. To intuitively understand this phenomenon, we present the experimental results in  Fig.~\ref{fig:downconverted-beacon-2}, where the \ubeacon advertises a continuous single-tone pulse at $40$ kHz and the \cbeacon transmits periodic chirps from $45$ kHz to $60$ kHz. Consequently, the receiver detects intact periodic chirps between $5\sim 15$ kHz. Our technical trick is that \emph{the spreading occurs in the air instead of at the transmitting side, and thereby no additional cost and bandwidth are required at \ubeacons}.

\begin{figure}[t!]
  \centering
  \includegraphics[width=0.9\linewidth]{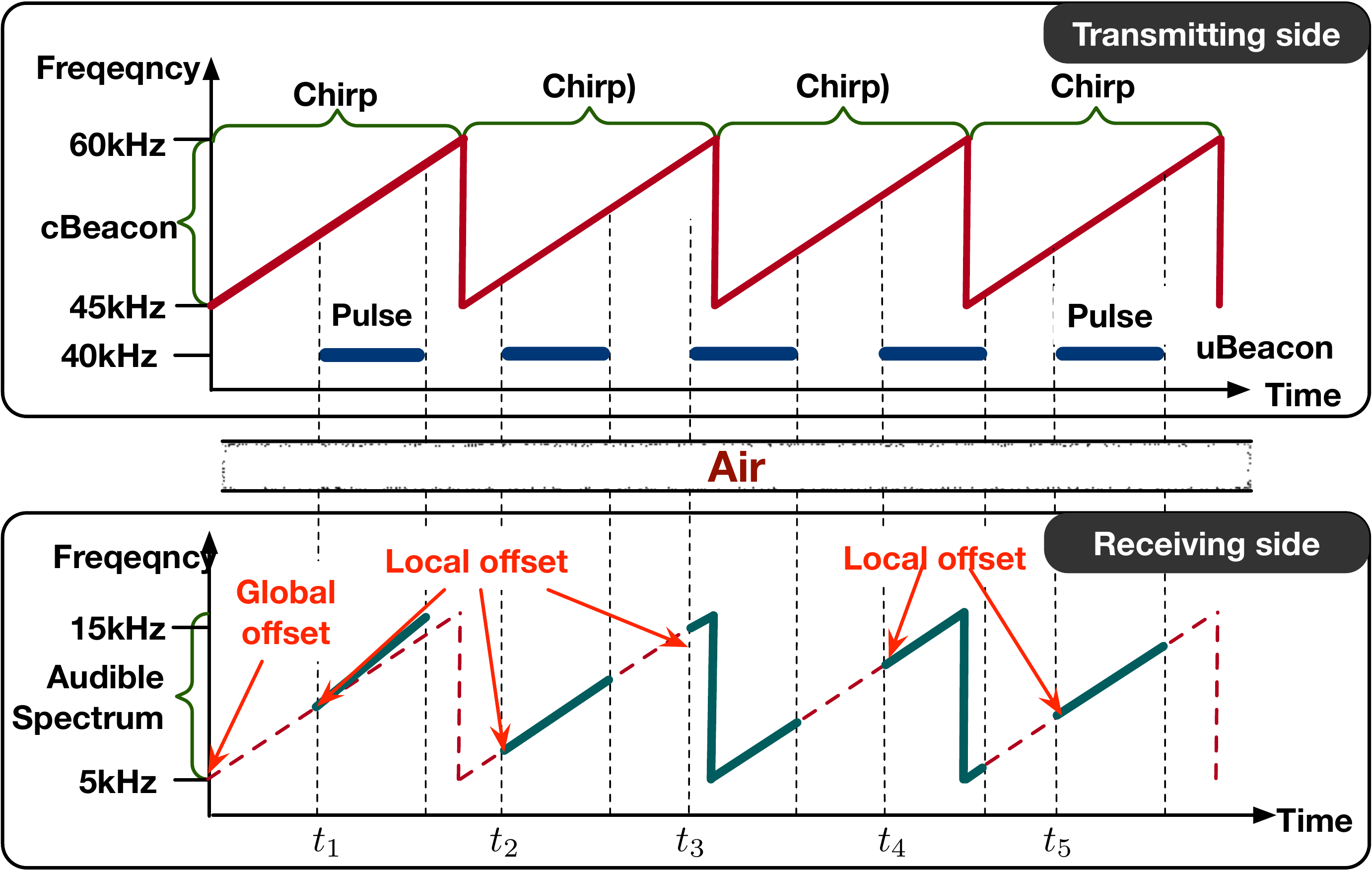}
    \vspace{-0.1cm}
   \caption{Detected dynamic chirp beacons. \textnormal{The \cbeacon transmits periodic chirps during $45\sim 60$ kHz, whereas the \ubeacon transmits a $300$ ms pulse beacon at $40$ kHz every $500$ ms. The receiver detects segments of the downconverted chirps.}}
  \label{fig:segmented-chirp}
    \vspace{-0.3cm}
\end{figure}

\textbf{Dynamic Chirp Spreading.} Now, let us consider what happens when \ubeacons periodically advertise \emph{pulse beacons}.  To illustrate our basic idea, we use the toy example in Fig.~\ref{fig:segmented-chirp}, where the \cbeacon transmits chirps continuously but a \ubeacon advertises short single-tone pulse beacons with spaces.
Downconversion occurs during the window when a pulse appears at the microphone; thus, the receiver can no longer detect an intact chirp but only segments of it.  In the figure, the green solid lines denote the captured chirp segments.  These segments are identical in time and bandwidth, both of which are determined by the pulse interval. However, their starting and ending frequencies may differ, \ie the spread chirps are \emph{dynamic} and depend on the propagation distance. We call this spreading technique \emph{Dynamic Chirp Spectrum Spread}  (DCSS). Fig.~\ref{fig:downconverted-beacon-3} shows the downconversion results under the same settings as shown in Fig.~\ref{fig:downconverted-beacon-2}, except that the signals of \ubeacon are changed to periodic pulses. The figure indicates that we can only observe the chirp segments where the unsampled parts are filled with dashed lines.

\subsection{Pinpointing Dynamic Chirp Beacons}

\begin{figure}[t!]
  \centering
  \includegraphics[width=\linewidth]{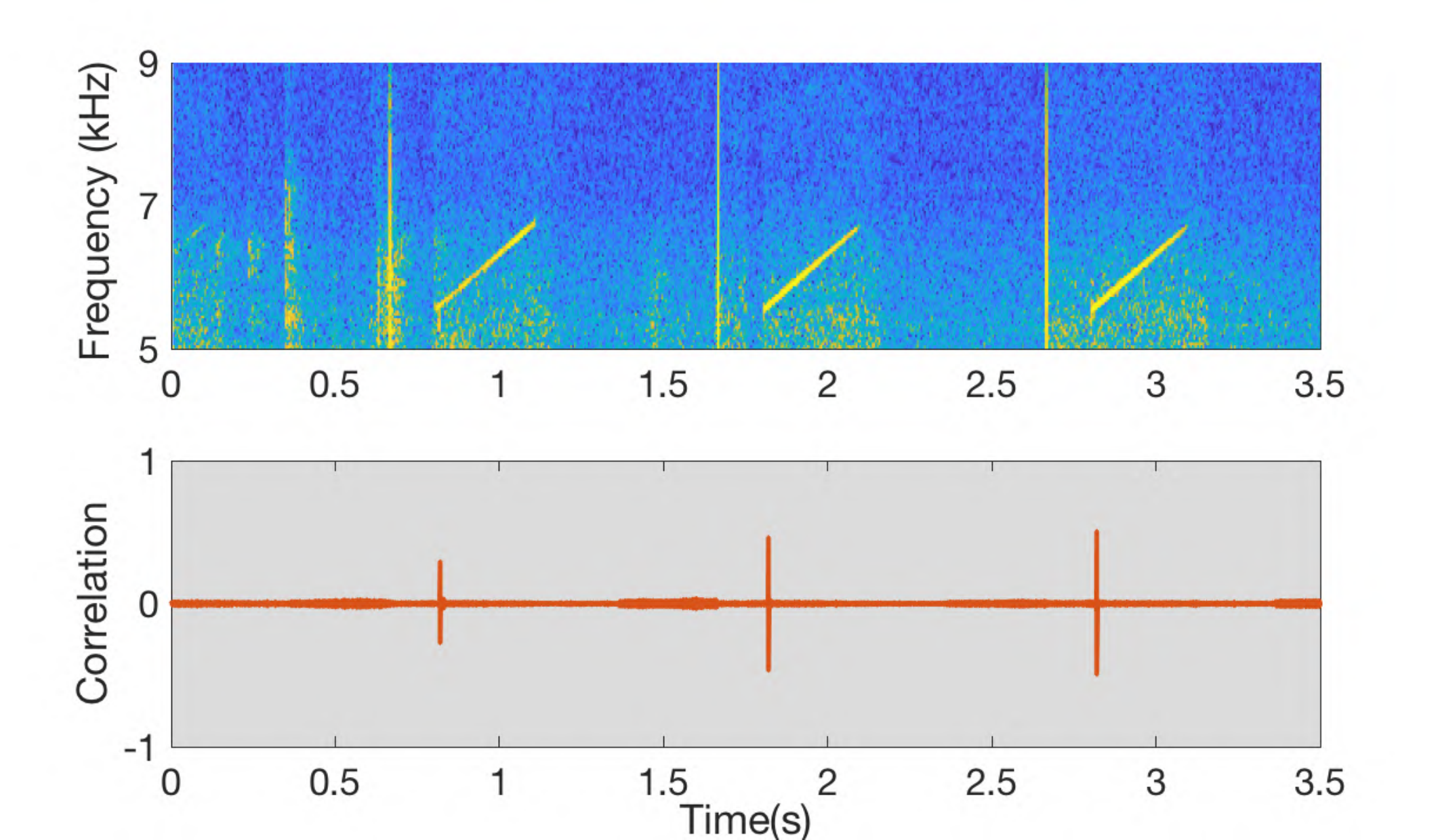}
    \vspace{-0.1cm}
   \caption{Correlation in the time domain. \textnormal{The above picture shows the spectrum via short-time Fourier Transform (STFT); the bottom picture shows the correlation result. The clear correlation peaks could be found exactly at the beginning of each chirp signal.}}
  \label{fig:correlation}
    \vspace{-0.3cm}
\end{figure} 

 The conventional correlation method which uses a static template, fails in identifying dynamic chirp segments in the present case because the template is not static. In particular, two unknowns, namely, the \emph{global offset} (denoted by $\Gamma$) and the \emph{local offset} (denoted by $\tau$), exist due to the dynamic condition. The global offset indicates the starting frequency of the \cbeacon's chirp when it first arrives at the microphone. The local offset indicates the starting frequency of the \ubeacon's pulse when it first arrives at the microphone.  These two types of offset are annotated in Fig.~\ref{fig:segmented-chirp}. Thus, ToA correlation must be performed in two dimensions as follows:
\begin{equation}\label{eqn:2d-correlation}\small
\begin{cases}
	(\Gamma, \tau) =  \mathlarger{\argmax}\limits_{(\Gamma, \tau)=(0, 0)}^{(K, N-K)} \frac{1}{K}\sum\limits_{k=\tau}^{\tau+K} \widetilde{S}[k] \cdot G(\Gamma, \tau,k)\\ 
	G(\Gamma, \tau,k)=\cos(2\pi (f_c-f_u +(\Gamma+0.5(\tau+k) \bmod K)\Delta f) t_k)
\end{cases}
\end{equation} 
where $\widetilde{S}$ is the received signal and $N$ is the total number of samples in the received signal. $N$ should be the double width of the \ubeacon's pulse to ensure that at least one chirp segment  is captured. $G$ is the \emph{dynamic chirp template}. The output of the objective function suggests a tuple, which enables the correlation to spike at the appropriate offset. Fig.~\ref{fig:correlation} illustrates the correlation results over the sample data shown in Fig.~\ref{fig:downconverted-beacon-3}.

\textbf{Optimization.} Solving the aforementioned objective function requires performing $N*K$ correlations, which will be a burdensome and energy-consuming task for a mobile device. We notice that once a smart device holds its position, all downconverted segments will share the global $\Gamma$, which is caused by the signal propagation delay from the \cbeacon. Thus, $\Gamma$ can be easily found by correlating an intact chirp template with the audio. The correlation spikes at $\Gamma$, which allows the chirp template to exactly cover all segments. Then, we start to slide $\tau$ to  determine the local offset. This optimization process can reduce the number of correlation to  $N+K$. Note that although the global offset is involved in the above equation, it is actually not used in the trilateration.

\section{Multiple Beacon Access}
\label{section:mba}
So far, we have discussed how \oursystem can estimate the ToA of a beacon out from  a single \ubeacon. This section discusses how \oursystem distinguishes beacons from multiple \ubeacons, \ie  \emph{multiple beacon access}.

\begin{figure}[t!]
  \centering
  \includegraphics[width=0.9\linewidth]{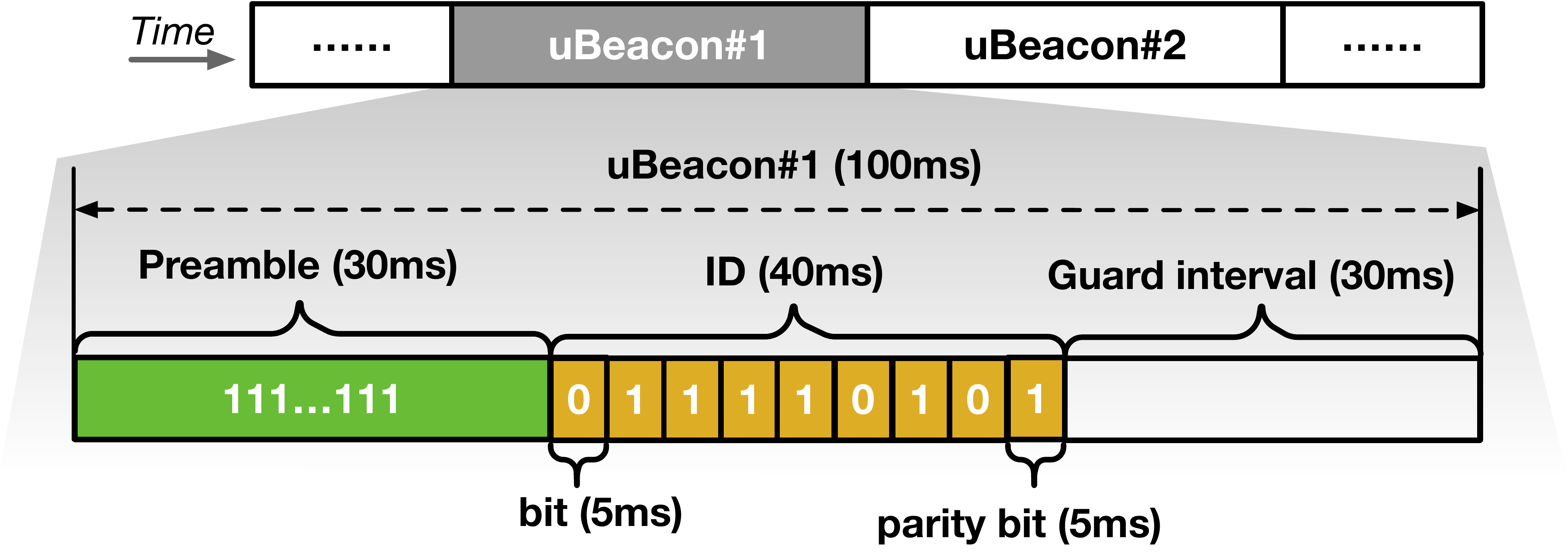}
    \vspace{-0.1cm}
   \caption{Multiple beacon access. \textnormal{Beacons are advertised in a predefined order to eliminate mutual interference. Each beacon contains an ID field to distinguish devices.}}
   \label{fig:format}
    \vspace{-0.3cm}
\end{figure}

\subsection{Time Synchronization}

Trilateration requires all reference devices (\ie \ubeacons) to be appropriately synchronized in time; otherwise,  each unsynchronized \ubeacon will introduce an unknown variable to the clock difference, thereby rendering Eqn.~\ref{eqn:pseudo-range} unsolvable. In this regard, we equip each device with a low-power WiFi chipset and adopt IEEE 1588 PTP for time synchronization. The classic Network Time Protocol (NTP) fails to satisfy our demand because it has an average delay of  $10$ ms, which leads to a ranging error of $10\,ms\times 340\,m/s=3.4\,m$ in \oursystem. By contrast, PTP can achieve a high accuracy up to $100\mu s$~\cite{TI-PTP} or produce a ranging error of $3.4$ mm. In \oursystem, the receiver does not need to be synchronized with \ubeacons because its clock offset is modeled in Eqn.~\ref{eqn:pseudo-range} already.

\subsection{Frequency/Time Multiple Access}

Previous works typically use FDMA to encode different beacons. However, this method does not work in our scenario because the downconverted beacon may sweep any segment of the chirp ( Fig.~\ref{fig:segmented-chirp}), which is highly correlated with the location of the receiver. To address this issue,  we adopt two-level encoding strategies in \oursystem. 

\begin{figure}[t]
  \centering
  \subfigure[Received audio signals]{
    \includegraphics[width=0.8\linewidth]{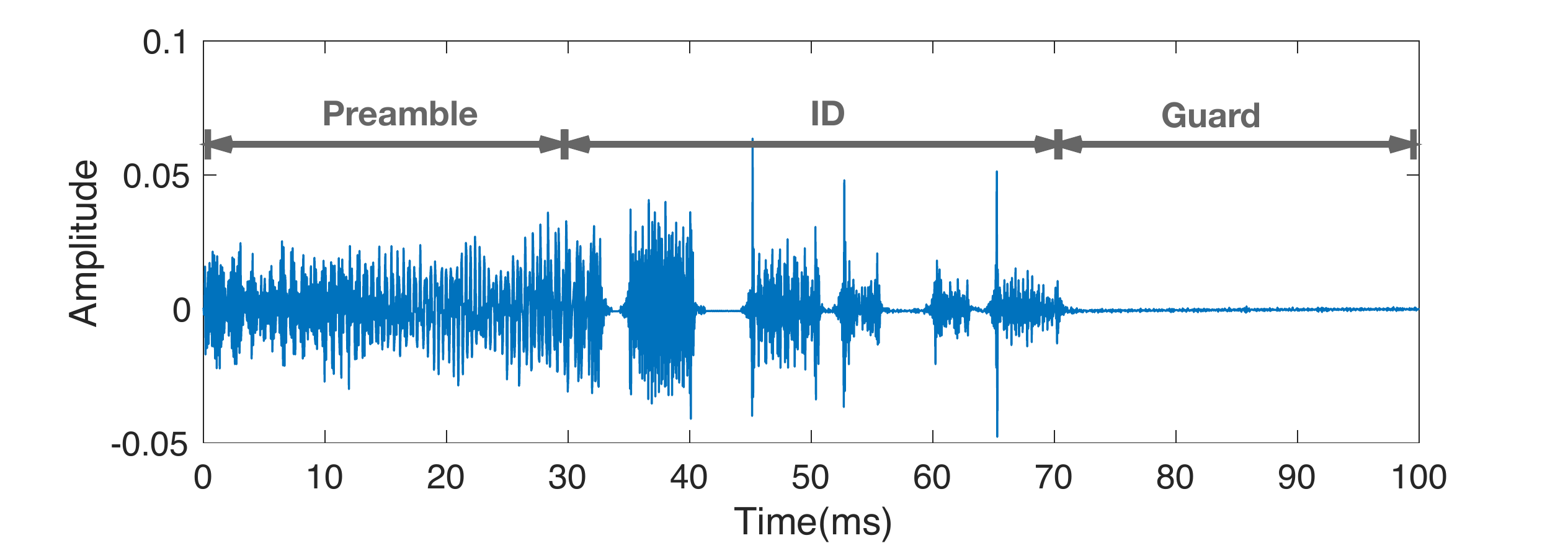}
    \label{fig:FM0-orginal-signal} 	
  }
  \subfigure[Decoded ID]{
    \includegraphics[width=0.8\linewidth]{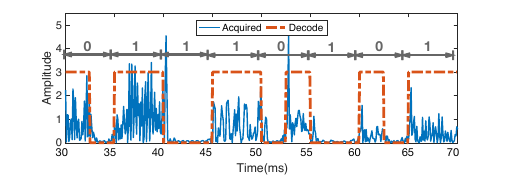}
     \label{fig:FM0-decode-signal}
  }
  \vspace{-0.2cm}
  \caption{Illustration of beacon decoding. \textnormal{(a) The whole beacon signals including the guard interval; and (b) zoom-in view of the ID field after the removal of the chirp carrier.}}
  \label{fig:decoding}
  \vspace{-0.3cm}
\end{figure}

\textbf{\cbeacon Encoding.} \oursystem assigns different \emph{sweeping slopes} to different \cbeacons, such that receivers can quickly locate itself at the room level.  Since a \cbeacon can cover about $50\times 50\,m^2$ area,  a single \cbeacon is enough for each room. Multiple \cbeacons are isolated through walls. We further apply on-off keying (OOK) to encode the IDs of \ubeacons covered by the same \cbeacon and schedule them in exclusive time slots. 

\textbf{uBeacon Encoding:}  Fig.~\ref{fig:format} shows the encoding of  pulse beacons. \ubeacons advertise their beacons in a predefined order to avoid mutual interference. Each beacon contains two fields: \emph{preamble} and \emph{ID}. The preamble is composed of $30$ ms pulses. The receiver uses the preamble to estimate ToA and to align at the ID field. The ID field contains $8$ bits, each of which has an interval of $5$ ms. These bits are encoded with FM0. Since FM0 requires all bits to be flipped in the beginning, it could avoid the emergence of a continuous $8$ bit pulses (\ie preamble-like ID). The last bit is reserved for parity checking. Thus, \oursystem can completely support $128$ \ubeacons that are covered by a single \cbeacon.  An additional $30$ ms guard blank is reserved at the end to avoid the interference from echoes. Different \ubeacons are scheduled to advertise in various slots. Users are allowed to configure the scheduling based on their practices, \eg allowing non-adjacent \ubeacons to advertise simultaneously to decrease delays; or increasing guard intervals for low-duty cycle.

\textbf{Beacon  Decoding}: The receiver initially seeks the beacon preamble in the recorded audio data via correlation ( see Eqn.~\ref{eqn:2d-correlation}). 
Once the preamble is obtained, the receivers can identify all the parameters of the chirp from the preamble. Then the audio data are multiplied with the chirp template (see Eqn.~\ref{eqn:beacon}) to remove the chirp-based carrier and acquire baseband signals over the ID field. Subsequently, `0' or `1' is decoded every $5$ ms by determining if a transition occurs during each bit interval (\ie Bi-Phase Space Coding, FM0). Finally, the receiver uses the decoded ID to determine the corresponding \ubeacon's location. Fig.~\ref{fig:decoding} illustrates an example of the received beacon signals, from which the ID is successfully decoded.

\textbf{ToA Estimation}. Suppose $M$ preambles are found in the audio data. Let $B_i$ denote the starting position of the $i^{th}$ preamble in the samples where $i=1,2,\dots, M$. $B_1$ is selected as the baseline and the $i^{th}$ beacon's ToA (\ie $t_i$) is calculated  as follows:
\begin{equation}
	t_i= \left((B_i-B_1)\bmod (100\,ms*f_s)\right)/f_s
\end{equation}
The $\bmod$ operation eliminates the scheduling delay, which is an integral multiple of $100$ ms, \ie \ubeacons are scheduled every $100$ ms. We also assume that beacon propagation does not traverse a scheduling period or beyond $100\,ms\times 340\,m/s = 34\,m$. This assumption is reasonable because a beacon propagating over $15$ m becomes nearly undetectable. The receiver substitutes $t_i$ into Eqn.~\ref{eqn:pseudo-range} for the trilateration.

 \section{Enhancement of Beacon}
 \label{section:enhance}
 
 The SNR of the second-order harmonics is weaker than the fundamental, which affects the accuracy and the effective range. To mitigate this issue,  we introduce multiple enhancement approaches at both transmitting and receiving sides. 

\begin{figure}[t!]
  \centering
  \includegraphics[width=0.8\linewidth]{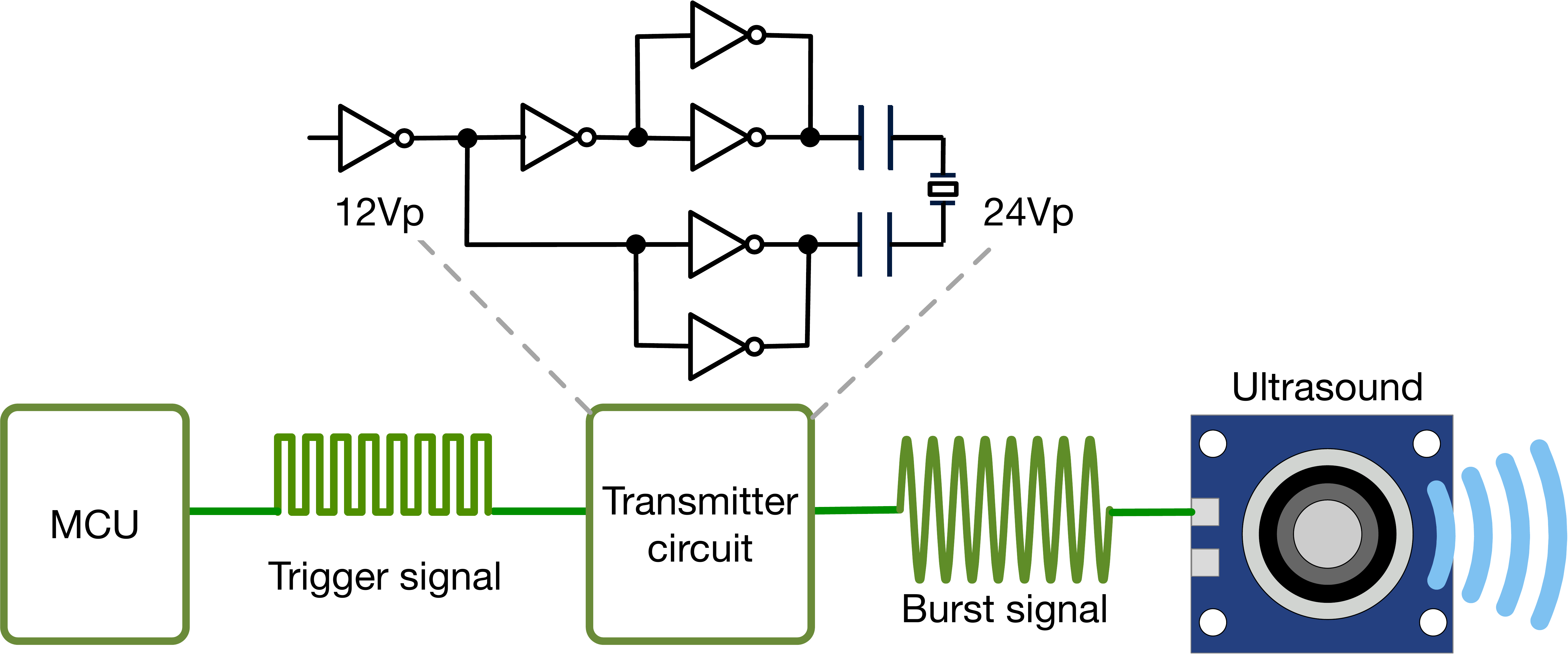}
   \caption{Transducer driver circuit}
   \label{fig:transducer-driver}
    \vspace{-0.5cm}
\end{figure} 
  
\subsection{Boosting Transmission}

\begin{figure}[tb]
  \centering
  \subfigure[Background noise]{
    \includegraphics[width=0.5\linewidth]{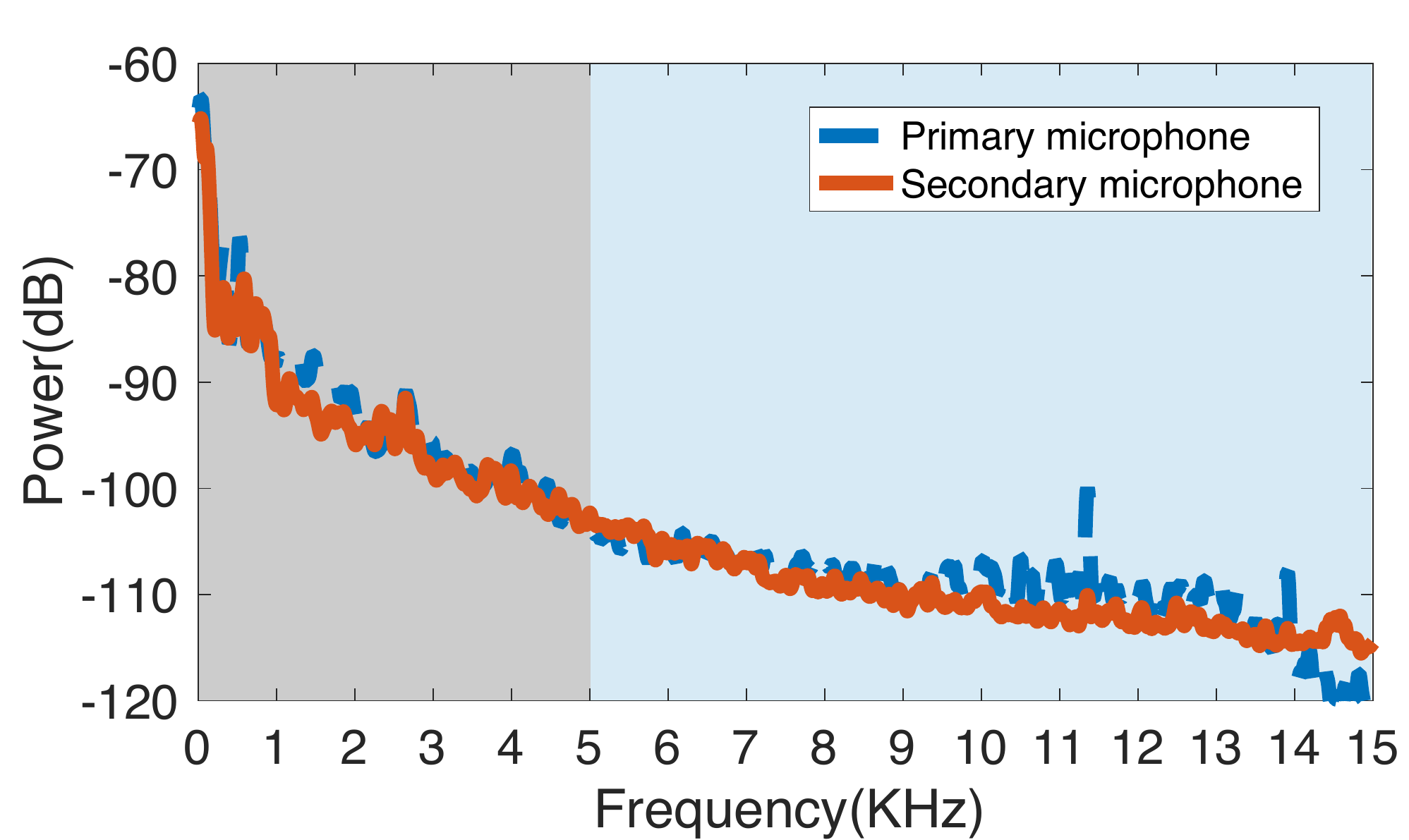}
    \label{fig:dual-noise} 	
  }%
  \subfigure[Downconverted chirps]{
    \includegraphics[width=0.5\linewidth]{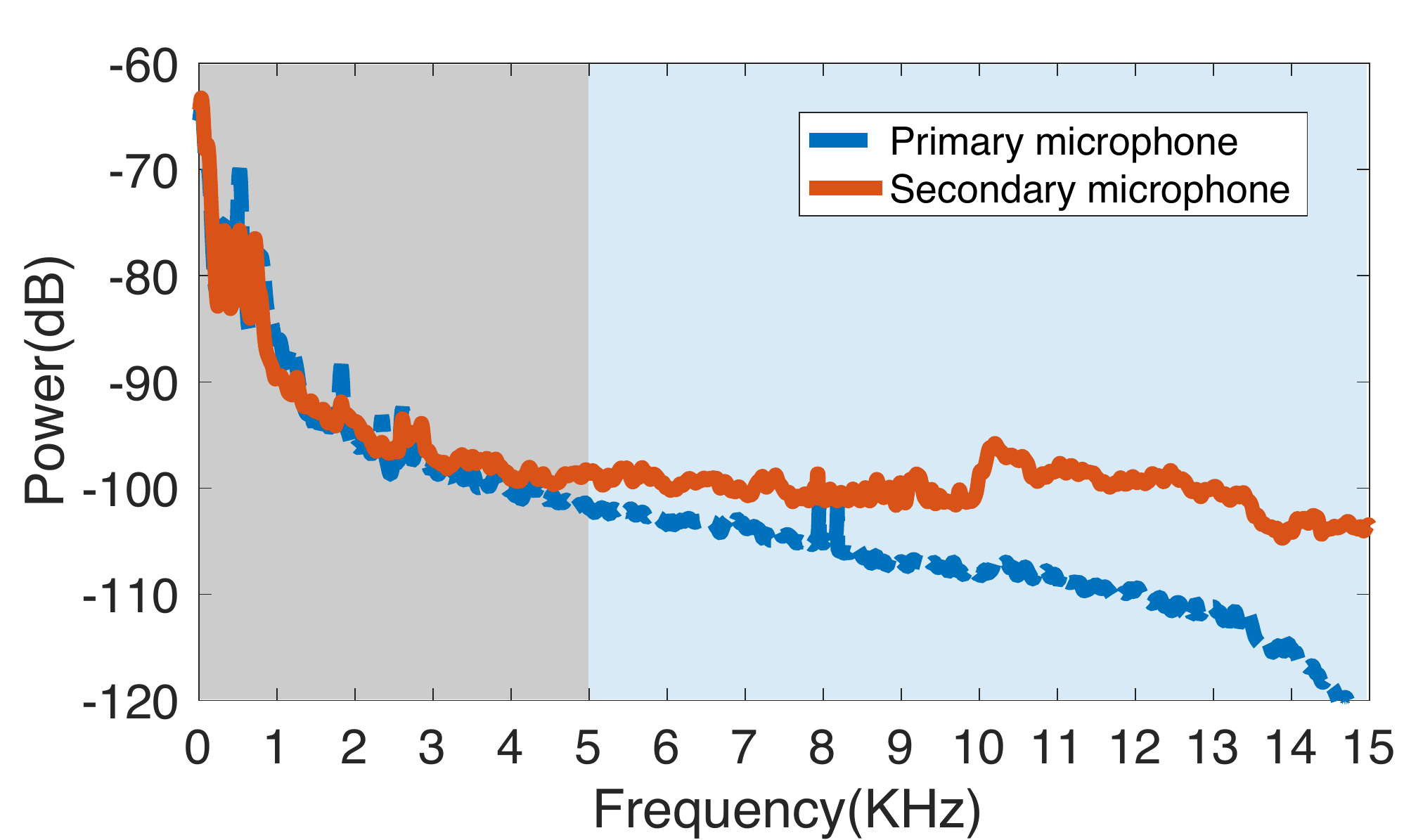}
     \label{fig:dual-chirps}
  }
  \vspace{-0.2cm}
  \caption{Audio PSD at two microphones. \textnormal{The operating spectrum of beacons is between $5\sim15$ kHz. (a) Two microphones record ambient noise with identical power levels; and (b) two microphones record similar chirps but with different power levels.}}
  \vspace{-0.5cm}
\end{figure}

 Firstly,  we boost the transmission at beacon devices.
   
 \textbf{Transmitter Circuit.} The conventional output circuit for the transducer is powered directly from the $5$ V supply, \ie standard input voltage.  We firstly use a boost converter to convert the $5$ V supply to $12$ V  DC voltage. Then,  we design a tricky driver circuit to raise and double the voltage. As Fig.~\ref{fig:transducer-driver} shows, we use an inverter gate to provide a $180^\circ$ phase-shifted signal to one arm of the driver. The other arm is driven by an in-phase signal. As a result, in spite of $12$ V power input, the current from two arms are constructively superimposed at the output. This design doubles the voltage swing at the output and provides about $24$ V peak voltage, which is almost twice than the conventional UPS.  
 
 \textbf{Transducer Array.} Using an array of transducers is a classical solution to improve the transmitting power of ultrasound. This solution has been adopted in many previous systems. For example, Cricket integrates $2$ transducers on a single beacon device, achieving a range of up to $10$m. Similar to ours, LipRead~\cite{roy2018inaudible} utilizes the second-order harmonic for \emph{long-range} voice attack.  Although an array consisting of  $61$ transducers is used in LipRead,  it only uses $5$ of them for each frequency segment and totally supports $6$ segments. Thus, it actually uses $5$ transducers for a specific frequency to achieve a maximum range of $30$ ft (or $9.1$ m).
 
 \textbf{Boost from \cbeacon}. The effective range of \oursystem is longer than LipRead because the strength of a downconverted beacon (\ie $A_{\downarrow}$)  depends not only on the transmitting power of the \ubeacon but also on that of the \cbeacon (see Eqn.~\ref{eqn:down-beacon}) . The transmitting power of the \cbeacon is $10,000\times$ stronger than \ubeacon, and thereby can enhance the strength of a beacon to a relatively higher level. With this in mind,  we integrate one transducer  in our \ubeacon prototype for energy and cost saving. However, it is easy to extend the current prototype to accommodate an array of transducers.

\subsection{Turbocharging Reception}
\label{section:turbocharging}

Secondly, we design an enhanced algorithm that operates at smart devices.  It attempts to draw beacons out of background noise when the receiver is far away from the beacon devices. We notice that  the majority of modern smart devices (\eg smartphones) have  two microphones.  The primary microphone is typically mounted on the bottom to ensure a direct acoustic path from the mouth.  The secondary microphone is mounted on top of the device to capture voice with a lower sound pressure level.  They are separated by approximately $10$ cm, and therefore, receive acoustic signals with different pressures. This condition offers an opportunity to turbocharge downconverted beacons. 

\begin{figure}[tb]
  \centering
  \subfigure[Before]{
    \includegraphics[width=0.5\linewidth]{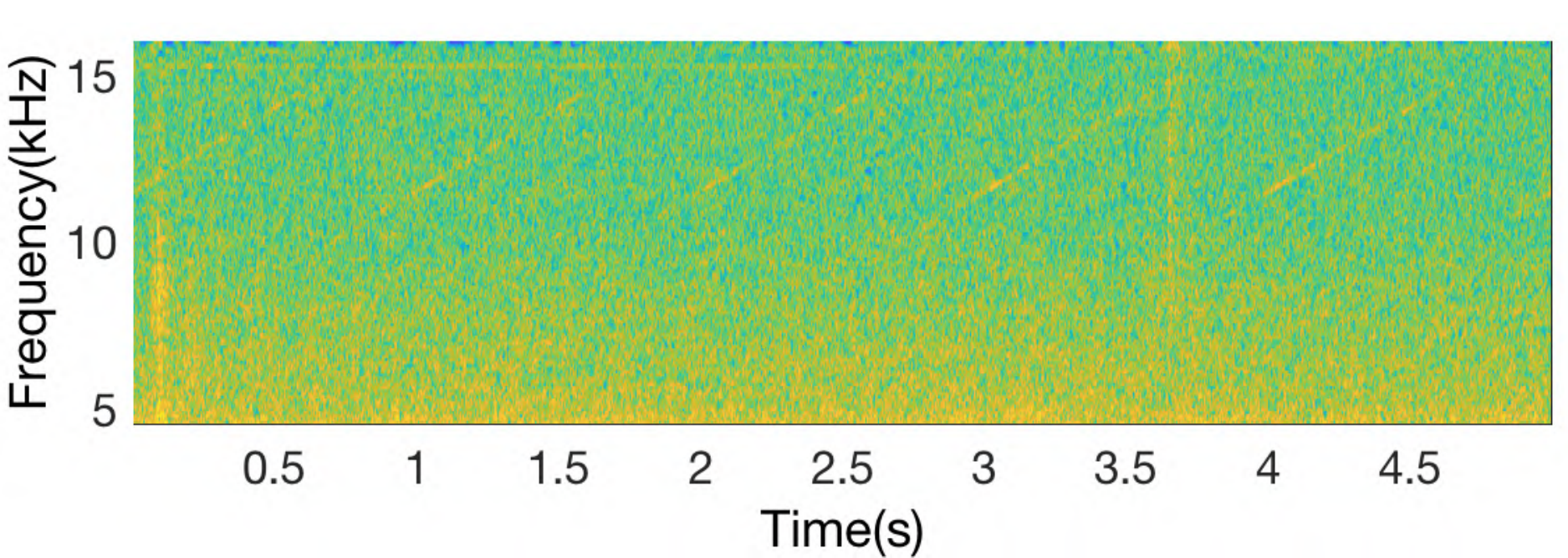}
    \label{fig:bench-environment} 	
  }%
  \subfigure[After]{
    \includegraphics[width=0.5\linewidth]{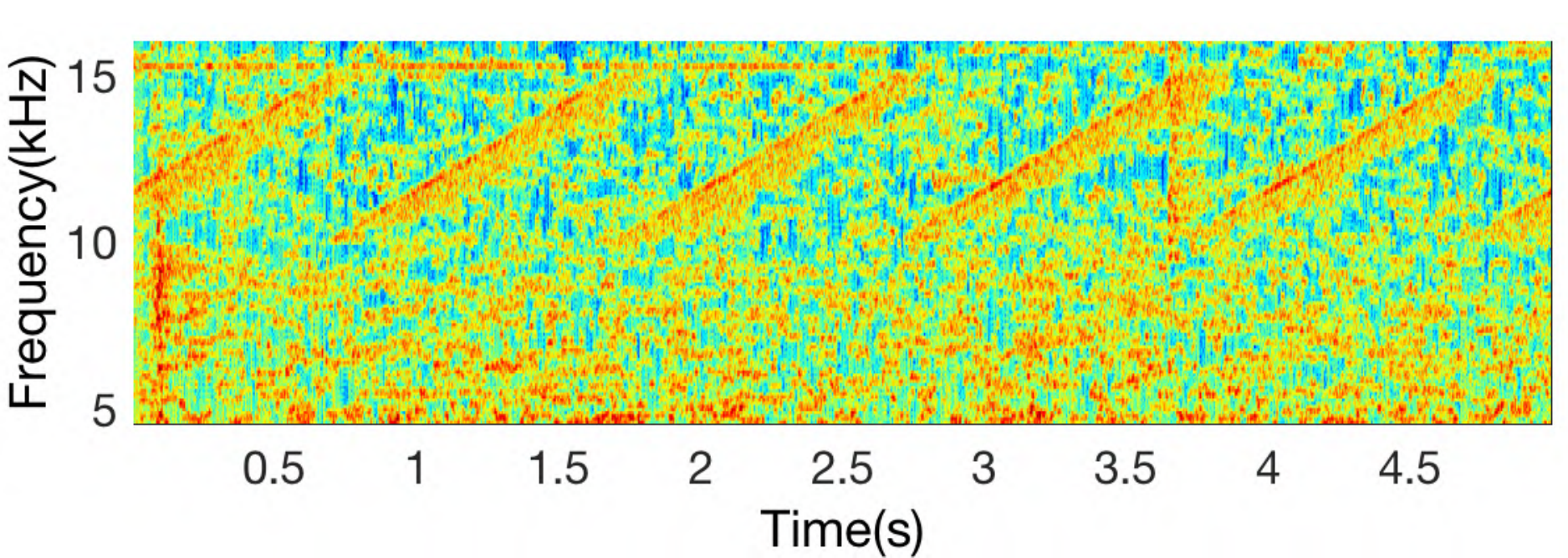}
     \label{fig:bench-energy}
  }
  \vspace{-0.2cm}
  \caption{Turbocharging results. \textnormal{The two figures show the beacon signals before and after turbocharging.}}
  \label{fig:turbocharging-example}
  \vspace{-0.5cm}
\end{figure}

\subsubsection{Understanding Two Microphones}

To understand the dual-microphone mode,  we conduct experiments to observe the responses of  two microphones installed in an iPhone X  as follows.

\textbf{Observation 1}. First, we use the two microphones to record ambient noise (without beacon signals)  for reference. The  power spectral densities (PSDs) of the  two audio data are shown in Fig.~\ref{fig:dual-noise}. The signals are nearly identical or homogeneous in terms of PSDs regardless of the locations of the two microphones. This phenomenon is also confirmed in our other experiments, which are conducted under noise-only conditions using the same mock-up phones but in a crowded place or a busy lecture room. Homogeneity is understandable. The  noise in our effective spectrum ($5\sim 15$ kHz) typically originates from thermal noise and household appliances.  These noises are ubiquitously present in a room without definitive directions.

\textbf{Observation 2.} Second,  we show the beacon-present PSDs  picked up by the two microphones  in Fig.~\ref{fig:dual-chirps}. The effective spectrum shows that the power level of the secondary microphone is considerably stronger than that of the primary microphone ($20$ dB higher) because our beacon devices are  mounted on the ceiling. Their signals must travel  additional centimeters to arrive at the primary microphone compared with the secondary microphone, thereby resulting in pressure difference. The directivity of ultrasound is stronger than that of audible sound due to its shorter wavelength. Such directivity further magnifies the difference.

\subsubsection{Turbocharging Algorithm}

The above two observations suggest that the PSDs of the two microphones are  quite similar regarding background noise, but become differentiable when beacons are present. These two key observations inspire us to design an enhancement algorithm at receivers. Intuitively, the downconverted beacons are chirp segments whose spectrums are dynamic, that is, the segment moves among different frequency bins. Moreover, the beacons are advertised every hundred of milliseconds. Both conditions provide idle windows to smart device to estimate the background noise. Specifically, when the PSD difference of two microphones is less than a threshold, the receiver starts to estimate the PSD of noise by assuming that no beacon signal is present currently; otherwise, the receiver estimates the PSD of signals and applies \emph{Wiener filter} to turbocharge the beacon signals. We omit the introduction of the filter due to space limitations and advise readers to refer to \cite{brown1992introduction,jeub2011robust} for details.  Fig.~\ref{fig:turbocharging-example} shows the effectiveness of the algorithm in an extreme case where the receiver is located $15$ m away from the \ubeacon. The figure shows that the turbocharged chirp signals become distinguishable although they are nearly drowned in background noise before filtering.

\begin{table*}[!t]
  \centering\small
  \caption{Comparison to Past Ultrasonic UPSs}
  \label{tab:motivation}
  \vspace{-0.3cm}
  \begin{threeparttable}
  \begin{tabular}{rrrrrrrccc}
    \toprule
     \textbf{Scheme}& \textbf{Type} &\textbf{Accuracy} & \textbf{Range}& \textbf{Cost/Node} & \textbf{Dim.}  & \textbf{Spectrum}&\textbf{Modulation}&\textbf{Inaudible}&\textbf{Compt.}\tnote{1} \\
    \toprule
    \textbf{Cricket~\cite{priyantha2000cricket}} & Transducer &$10$ cm & $10$ m &$10$ \$  & $2$D &$40$ kHz & Pulse &Yes& No \\
    \textbf{Dolphin~\cite{hazas2002novel}} &Transducer&$2.34$ cm &$3$ m&$10$ \$&$3$D&$20\sim 100$ kHz&DSSS&Yes & No\\
    \textbf{PC~\cite{lazik2012indoor}} & Speaker&$4.3$ cm & $50$ m & $50$ \$ & $3$D &$19\sim 23$ kHz & Chirp &No & Yes\\
\textbf{BeepBeep~\cite{peng2007beepbeep}} & Speaker&$ 0.8$ cm & $5$ m  & -- & $1$D & $2-6$ kHz & Chirp&No & Yes \\
  \textbf{ApneaApp~\cite{nandakumar2015contactless}}&Speaker&$2$ cm &$1$ m &--&$1$D&$18-20$ kHz&Chirp&No & Yes\\
    \textbf{\texttt{ALPS~\cite{lazik2015alps}}} & Transducer &$ 16.1$ cm &$40$ m &$>10$ \$ & $3$D&$20\sim  21.5$ kHz&  Chirp&No& No\\
    \textbf{\texttt{TUPS}} & Transducer &$3.51$ cm &$8$ m & $10$ \$ & $3$D&$40$ kHz&  Pulse&No& No\\
    \textbf{\oursystem} & Hybrid &$4.95$ cm & $6$ m & $5$ \$\tnote{2} & $3$D&$40\sim  65$ kHz&  DCSS&Yes& Yes\\
    \hline
    \textbf{Tagoram~\cite{yang2014tagoram}}
     & RFID &$8$ mm & $12$ m & $0.1$ \$ & $3$D&$820$ MHz&  ASK  & No & No\\
    \textbf{ArrayTrack~\cite{xiong2012arraytrack}} & WiFi &$57$ cm & $100$ m & -- & $3$D&$2.4$ GHz&  FDMA  & No & No\\
    \textbf{iBeacon~\cite{estimote}} & Bluetooth &$1$ m & $150$ m & $25$ \$ & $1$D&$2.4$ GHz&  DCSS  & No & Yes\\
    \textbf{WiTrack~\cite{adib20143d}} & -- &$13$ cm & -- & -- & $3$D&$5.46\sim 7.25$ GHz&  FMCW  & No & No\\
    \bottomrule
  \end{tabular}
  \begin{tablenotes}
      \small
      {\item[1] The column of `Compt.' indicates if the solution is compatible with today's smart devices.}
      {\item[2] The cost in \oursystem does not contain the price of \cbeacon, each of which cost about 60\$ and is shared by multiple \ubeacons.}
    \end{tablenotes}
  \end{threeparttable}
\end{table*}

\section{Implementation}
\label{section:implementation}

This section presents the implementation of two types of beacon devices. Their prototypes are shown in Fig.~\ref{fig:devices}.

$\bullet$ \textbf{\ubeacons}. The \ubeacons have functions similar to those of devices designed in Cricket~\cite{priyantha2000cricket}. A simple solution is to directly apply their designs to our system. However, the Cricket design was released ten years ago and many of its components are already  outdated. Thus, we have to redesign our own hardware.  The core function is to drive an on-board ultrasonic transducer (MA40S4S~\cite{transducer-datasheet}) to speak at $40$ kHz at a specific time point.  We also reserve an ultrasonic receiver for testing, which is not required for \oursystem.  The low cost WiFi chip   ESP32~\cite{esp32} from Espressif systems is integrated for the PTP protocol and online configuration. We synchronize \ubeacons every $32 s$ for energy saving where the beacon broadcast is advised between $0.25 s\sim 64 s$. Each \ubeacon is supplied by two $5$ V-batteries.  

$\bullet$ \textbf{\cbeacon}. We use a general vector signal generator to produce periodic ultrasonic chirps, which are further transmitted to the air through an ultrasonic speaker (Vifa~\cite{vifa}). The output power is tuned to $10$ W. It sweeps the  $10$ kHz band from $45$ kHz to $55$ kHz over $100\,ms$, \ie $100$ kHz per second. As a result, the downconverted spectrum is between $5\sim 15$ kHz regarding to the $40$ kHz of \ubeacons. We intentionally  reserve the top $7$ kHz bandwidth (\ie $15\sim 22$ kHz) to avoid the signal distortion due to the non-ideal transition of the low-pass filters at microphones.   

Note that the above spectrum is only given as an example setting. Some pets (\eg dogs) might be still sensitive to the $40\sim 65$ kHz. We could raise the spectrum to a higher range (\eg $70\sim 80$ kHz vs. $65$ kHz) once the relative downconverted spectrum falls into $5\sim 15$ kHz. In addition, our chirp signals are repeated every $70$ ms. Such repetition could introduce a $14$ Hz sound. However, $14$ Hz sound belongs to subsonic wave and is far below than the lower bound of the human hearing system (\ie $50$ Hz). Thus, users are not aware of its presence.

\section{Evaluation}
\label{section:evaluation}

A total of $15$ \ubeacons are deployed in a meeting room with an area of $9\times 3 m^2$ in our department. The speed of sound is set to $344.38$ m/s in our experiments. The speed of sound is relevant to the temperature, thus a calibration before the experiments is performed. To test smart devices as many as possible and in a cross-platform manner, we directly use a built-in recorder of a mobile OS (\eg iOS or Android) to record audio clips for post-processing in MATLAB.  We use \ubeacons equipped with a single transducer by default and an iPhone X as the default receiver unless noted.

\subsection{Evaluation in Zero-Dimension}

We start by qualitatively comparing \oursystem with state-of-the-art UPSs and evaluate the accuracy of ToA estimation in zero-dimension.

\subsubsection{Comparison with State-of-the-Art}
We compare  \oursystem with past UPSs from nine different perspectives, as listed in Table ~\ref{tab:motivation}. (1)  Only a few UPSs use ultrasonic speakers as anchors due to their high prices. PC is an attempt in this direction. It achieves a good accuracy at an extremely higher cost. (2) Unlike normal UPSs, Dolphin deploys custom-made ultrasonic receivers to locate  transmitters, but has a shorter range (\ie $3$ m);  (3) ALPS designs an embedded ultrasonic speaker, which can achieve the operating range of $40$ m.  Due to the lack of hardware, we present the mean accuracy in the table only. (4) both BeepBeep and ApneaApp use the speakers in the smart devices for ranging. (5) TUPS is the UPS that we implemented across \ubeacons but operates at the fundamental frequency. TUPS orientates to ultrasonic receivers, being similar to the ALPS, Dolphin and Cricket. Thus, we choose TUPS as a benchmark baseline in our evaluation.
Particularly, we have the following observations:  (a) Previous UPSs use either ultrasonic transducers or speakers, whereas \oursystem is a unique system that uses the two components jointly to build a hybrid UPS; (b) \oursystem achieves comparable accuracy and effective range as transducer-built UPSs, and meanwhile the unit cost is maintained at an acceptable level; (c) \oursystem is the unique system that operates at the ultrasonic spectrum, but remains compatible with current  smart devices.

We also list the results of other four typical RF-based solutions for comparison. Usually, RF-based solutions have longer operating range (\eg around $100$ m), but their accuracies are limited to  meter or sub-meter level. RFID-enabled Tagoram behaves a good accuracy but requires the prior knowledge of track. Importantly, UHF RFIDs are unavailable by smart devices due to lack of UHF readers. Bluetooth-enabled iBeacon might be the most widely used commercial localization technology, which was initiated by Apple Corp. Our real experiments show that the average accuracy of an iBeacon is around one meter.  WiTrack locates targets at cm-level using RF reflections at the cost of almost 2 GHz bandwidth.  Compared against RF solutions, our work  offers a trade-off solution between the practicality and the accuracy to the currently available smart devices and to meet the growing demand on indoor high precision localization.

\subsubsection{Accuracy in ToA Estimation} 

\begin{figure*}[t!]
  \centering
  \begin{minipage}{0.33\linewidth}
  	\centering
    \includegraphics[width=\linewidth,height=2.3cm]{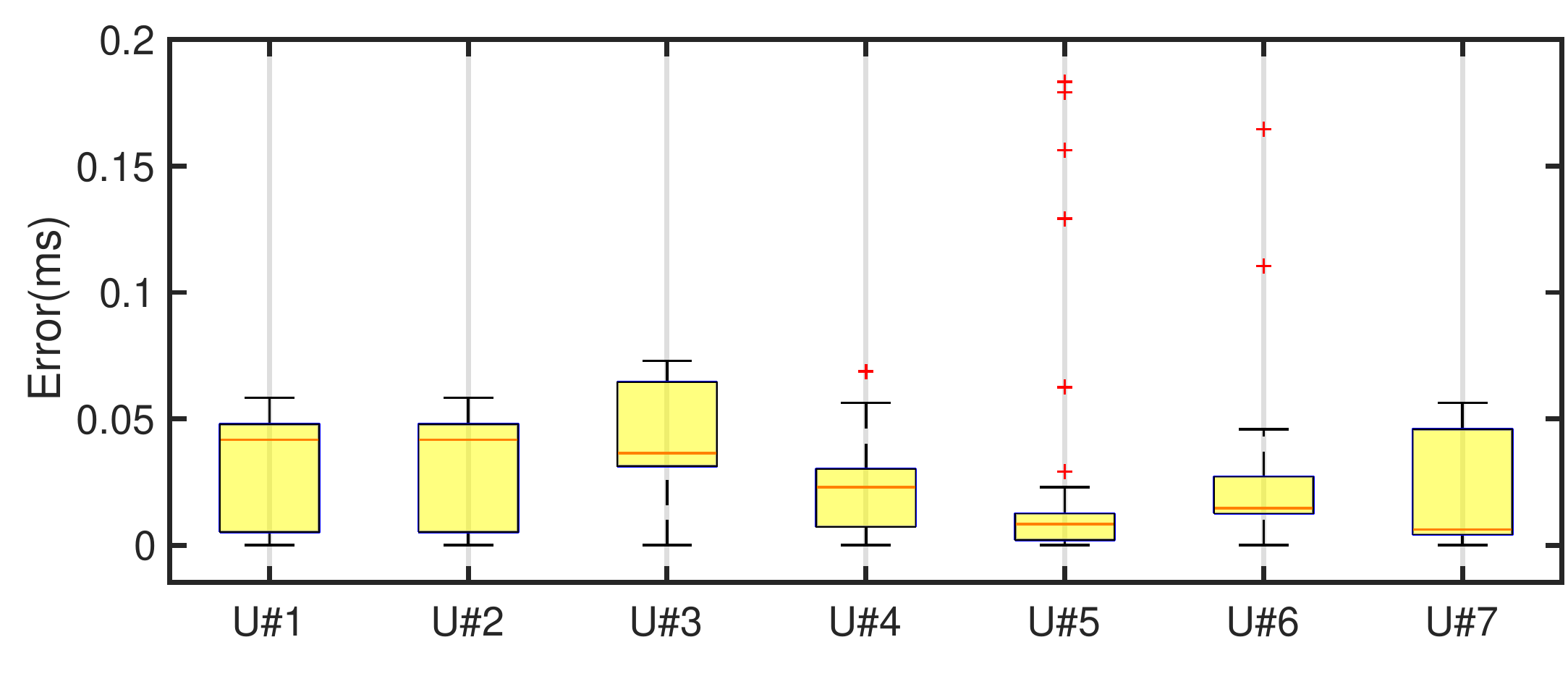}
   \caption{ToA estimation}
   \label{fig:accuracy-toa}
 \end{minipage}%
 \begin{minipage}{0.33\linewidth}
  	\centering
  	\includegraphics[width=\linewidth,height=2.3cm]{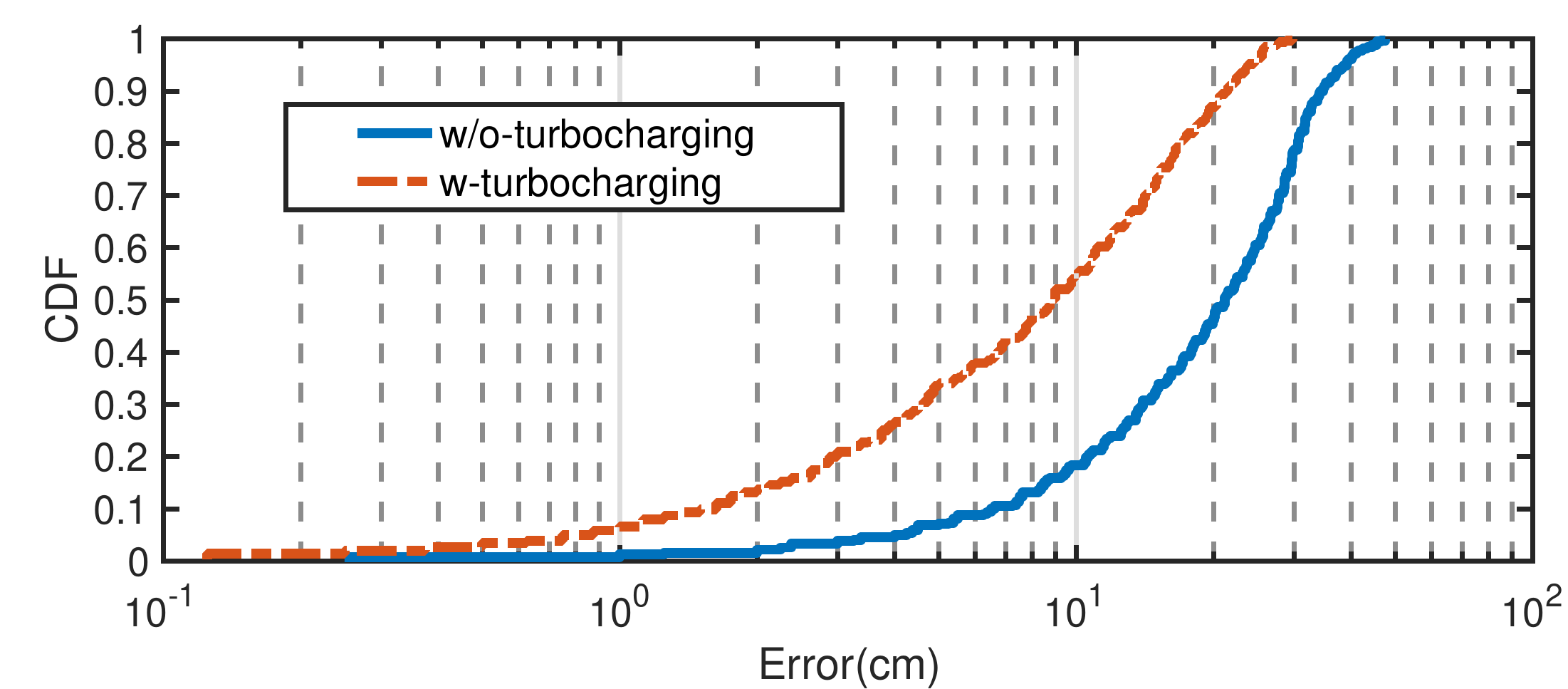}
   	\caption{Turbocharging}
   	\label{fig:bench_1d_turbocharging}
\end{minipage}%
\begin{minipage}{0.33\linewidth}
  \centering
  \includegraphics[width=\linewidth,height=2.3cm]{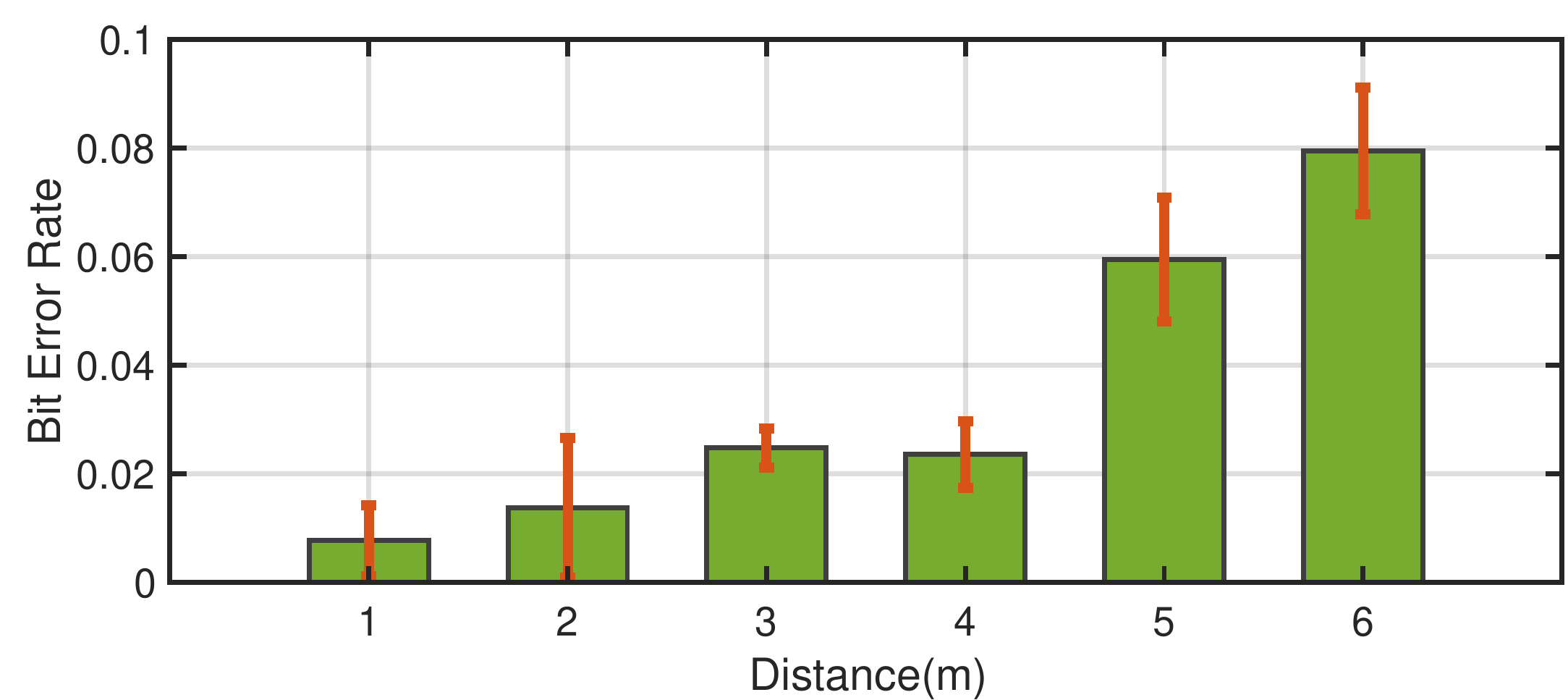}
   \caption{Bit error rate}
   \label{fig:2d-ber}
  \end{minipage}
\end{figure*}

As ToA estimation is the foundation of the trilateration, we evaluate the accuracy of ToA estimation. In the experiment,  we place seven different \ubeacons $2$ m away from the receiver while holding the receiver at a static location (\ie in zero-dimension). The \ubeacons advertises beacons every $5$ s. We record a $10$ min audio across these \ubeacons. Given that the distances between the receiver and the \ubeacons remain unchanged at all times, the downconverted beacons should arrive exactly every $5$ s. We compute the ToA of the remaining beacons by considering the ToA of the first beacon as the reference. These relative times should be an integral multiple of $5$ s.  Fig.~\ref{fig:accuracy-toa} shows the errors of the estimated ToA across the seven \ubeacons. 

The result suggests that approximately $0.03$ ms error occurs in the estimation even if the receiver remains at its location, mainly due to the synchronization time delay and the additional time consumption for processing audio data at the receiver, \eg moving data from an audio capture system to a memory or disk. Such estimation delay will incur a potential $0.03\,ms\times 344.38\,m/s=1.03\,cm$ error in the subsequent ranging or localization results. Nevertheless, such error is rather stable because the maximum standard deviation is around $0.02$ ms. This experiment demonstrates the feasibility of using downconverted beacons for localization.

\subsection{Evaluation in One-Dimension}

Then, we  evaluate the ability of \oursystem in one-dimension, that is, both the \ubeacon and the \cbeacon are fixed at their positions. The receiver is moved away from the \ubeacon by following a straight line. Their distance is increased from $40\,cm$ to $800\,cm$ with a step interval of $10cm$.  We are interested in determining \oursystem's performance as a function of distance.

\begin{figure*}[t!]
  \centering
  \includegraphics[width=0.95\linewidth]{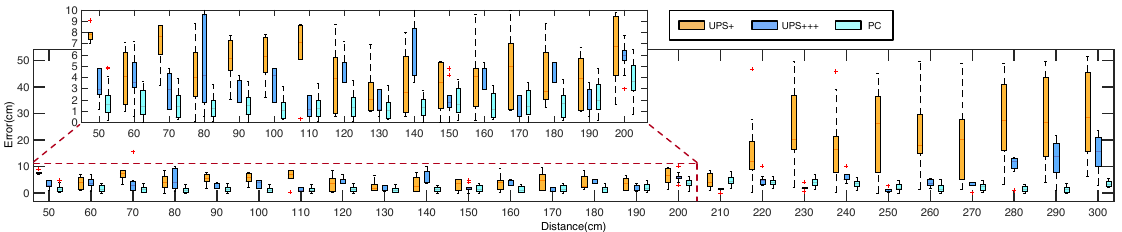}
    \vspace{-0.2cm}
   \caption{Ranging accuracy as a function of distance. \textnormal{UPS+ integrates a single transducer on a \ubeacon; UPS+++ integrates three transducers on a \ubeacon; PC refers to \cite{lazik2012indoor}.}}
  \label{fig:1d-vs-distance}
\end{figure*} 

\subsubsection{Accuracy in Ranging}

We initially investigate \oursystem's ranging accuracy. We compute the pseudo-ranges when the receiver is located at different positions. Adopting the pseudo-range at $40$ cm as the reference, the displacements relative to the reference in other positions are calculated.  The experimental trials are conducted $20$ times in each position. The  \emph{effective range} is defined as the maximum distance when the median ranging error is less than $20$ cm. The ranging accuracy is shown in Fig.~\ref{fig:1d-vs-distance}, which only shows the results when the distance is less than $300$ cm due to the space limit. We have the following findings:

$\bullet$ \textbf{Single-transducer made \ubeacon}: Firstly, we equip a single transducer on the \ubeacon.  \oursystem achieves a mean accuracy of $4.85$ cm and a standard deviation of $2.8$ cm when the distance is less than $2.5$ m. However, the mean error increases to $22.61$ cm when the distance is beyond $2.5$ m. The effective range is about $3$ m. This is due to the facts:  the transmitting power of a \ubeacon is limited to $0.9$ mW, and the adoption of second-order signals suffers from a larger attenuation compared with that of the first-order signals. Both factors decrease the SNR of downconverted beacons, thereby reducing the accuracy, when the receiver is away from the \ubeacon.

$\bullet$ \textbf{Three-transducer made \ubeacon}: Secondly, we equip three transducers (\ie \texttt{UPS+++}) on the \ubeacon to form a simple ultrasonic array.  \texttt{UPS+++} has a mean accuracy of $3.6$ cm when the distance is less than $3$ m. Correspondingly, the effective range is extended to $6$ m. This result almost reaches the range of TUPS, which operates at the first-order signal but is equipped with a single transducer.  The cost is only increased $3$ \$ than TUPS. This suggests that multiple transducers can indeed enhance the transmitting power of ultrasound. 

$\bullet$  \textbf{Speaker made beacon device:} Thirdly,  another simple approach for lengthening the ranging distance is to increase the transmitting power similar to PC, which uses the ultrasonic speaker to transmit chirps at $19-23$ kHz. The transmitting power is tuned at $10$ W, which is $10,000\times$ that of our \ubeacon. Consequently, the ranging accuracy remains at $4.3$ cm even if the receiver is located at a distance of $6$ m.

In summary, \oursystem can achieve centimeter-level ranging accuracy and works comparably with ultrasonic ranging within an effective range of approximately $3$ m, when \ubeacons are equipped with single-transducers. We believe that $3$ m range can fulfill major demands in practice, particularly when \ubeacons are  attached to ceilings or walls. Of course, we can increase the ranging distance by enhancing the transmitting power or integrating multiple transducers if necessary.

\subsubsection{Accuracy after Turbocharging}

We evaluate the effect of the PSD-aware turbocharging algorithm at a smart receiver equipped with two microphones.  Android smart-phones are allowed to select audios through two microphones by turning on the stereo mode. We place a Huawei smart phone at a distance of $5$ m. Fig.~\ref{fig:bench_1d_turbocharging} shows the comparisons of ranging accuracy with and without turbocharging.  The median error \emph{without} turbocharging is approximately $21.13$ cm due to the confusion from background noise. Intuitively, a $50$-sample shift will incur a $0.1\,ms$ time shift or $27\,cm$ error in ranging. Dozens of samples have been obscured in this case. 
To improve SNR, we apply the turbocharging algorithm to ToA estimation. Consequently, the median error rapidly drops below $9$ cm, which outperforms the result twice. This is because the turbocharging can improve the energy of the  correlation peak by thrice more than that without turbocharging.

\begin{figure*}[t!]
  \centering
  \begin{minipage}{0.33\linewidth}
  	\centering
  	\includegraphics[width=\linewidth,height=2.3cm]{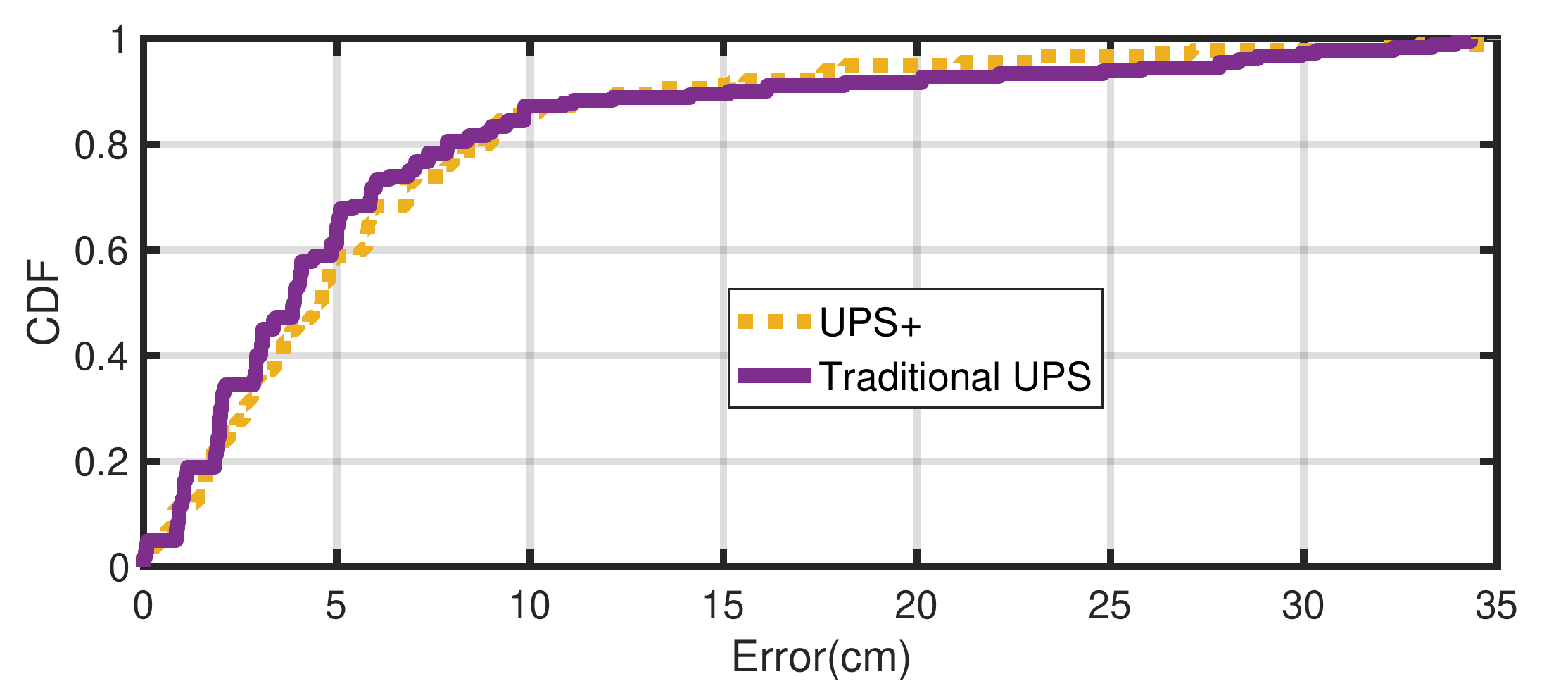}
    \caption{Localization accuracy}
    \label{fig:2d-comparison}
 \end{minipage}%
 \begin{minipage}{0.33\linewidth}
  	\centering
  	\includegraphics[width=\linewidth,height=2.3cm]{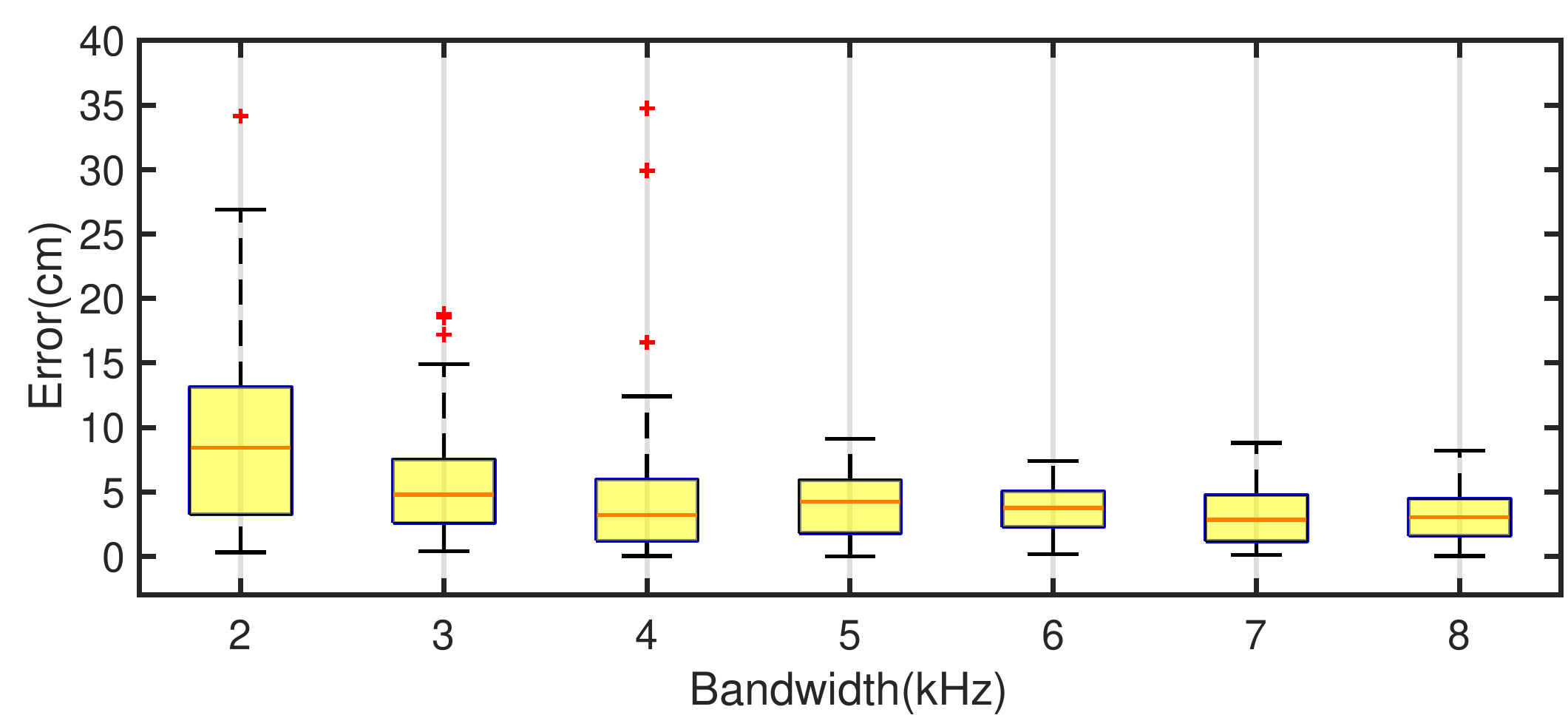}
   \caption{Error vs. bandwidth}
   \label{fig:2d-bandwidth}
\end{minipage}%
\begin{minipage}{0.33\linewidth}
  \centering
  \includegraphics[width=\linewidth,height=2.3cm]{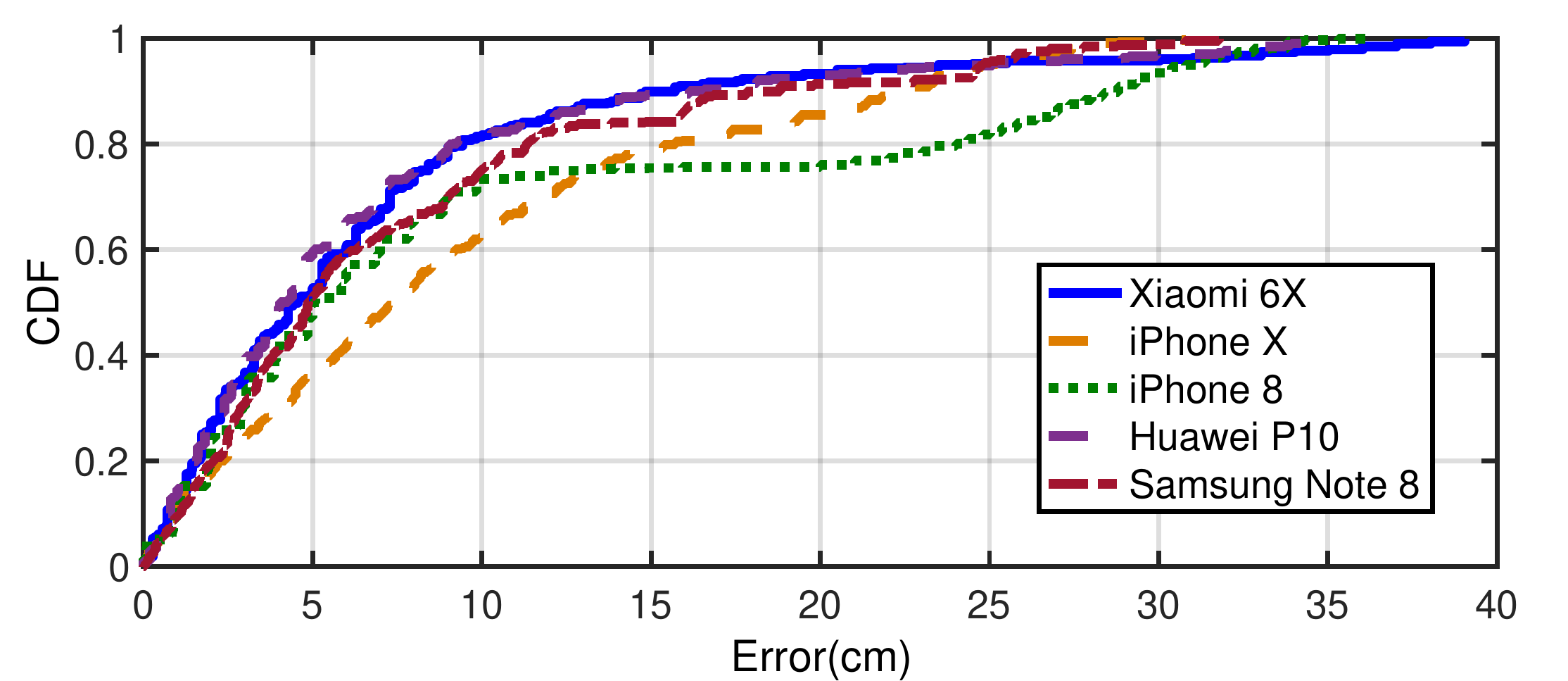}
   \caption{Accuracy vs. Device}
   \label{fig:2d-device}
  \end{minipage}
\end{figure*}

\begin{figure*}[t!]
  \centering
  \begin{minipage}{0.33\linewidth}
  	\centering
  	\includegraphics[width=\linewidth,height=2.3cm]{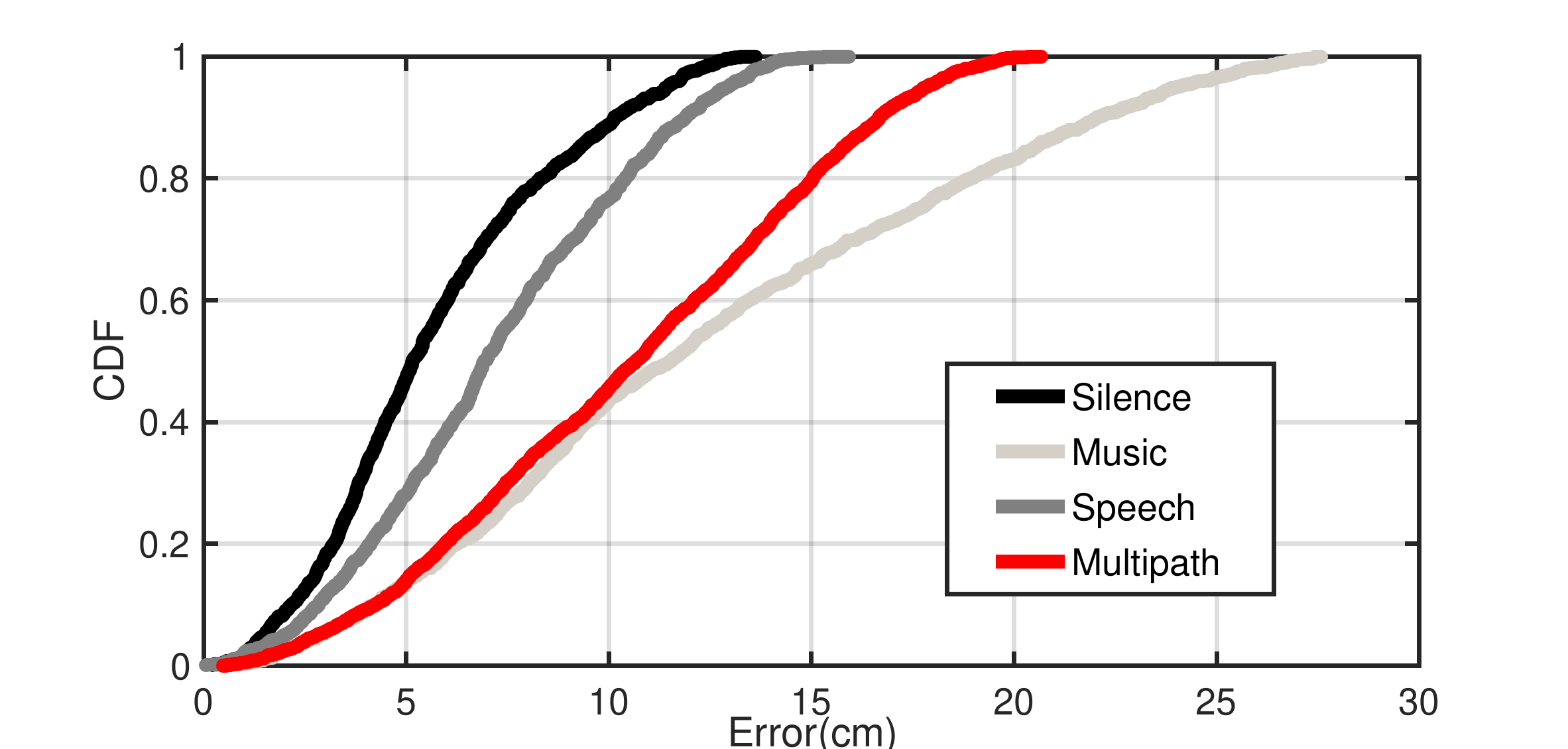}
    \caption{Accuracy vs. environment}
    \label{fig:bench-environment}
 \end{minipage}%
 \begin{minipage}{0.33\linewidth}
  	\centering
  	\includegraphics[width=\linewidth,height=2.3cm]{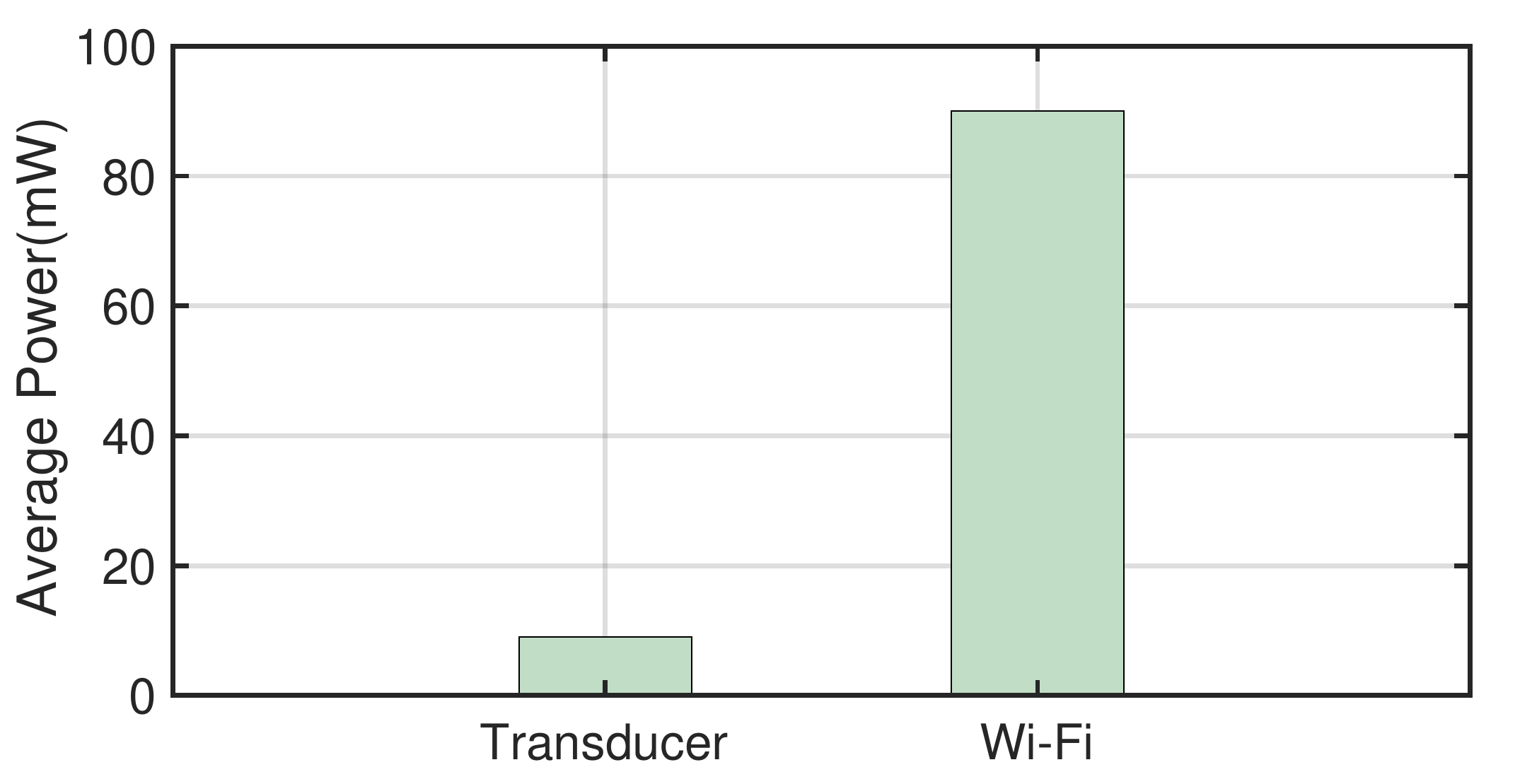}
   \caption{Energy vs. components}
   \label{fig:bench-energy-comp}
\end{minipage}%
\begin{minipage}{0.33\linewidth}
  \centering
  \includegraphics[width=\linewidth,height=2.3cm]{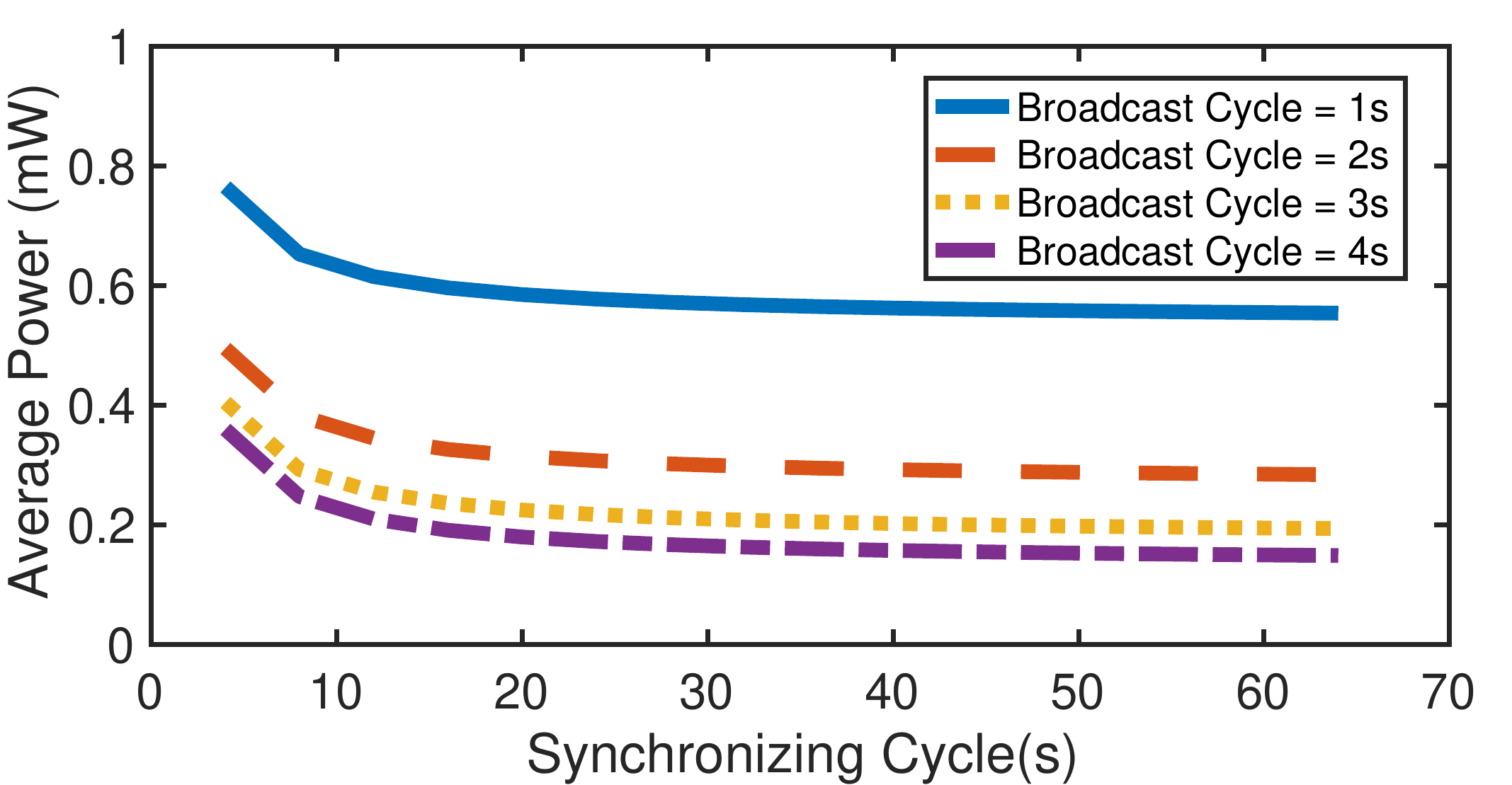}
   \caption{Energy vs. cycle}
   \label{fig:bench-energy-cycles}
  \end{minipage}
\end{figure*} 

\subsubsection{Accuracy in Decoding}

 Subsequently, we evaluate the \oursystem's decoding ability. We skip the identification test of the \cbeacon because the results are always accurate regardless of how the \cbeacon's sweeping slope is changed. The main reason for this condition is that transmitting signals of the \cbeacon is relatively strong ($10,000\times$ that of the  \ubeacons), which results in easy decoding. Here, we only focus on the decoding of \ubeacons' IDs. Fig.~\ref{fig:2d-ber} shows bit error rate (BER) as a function of the distance. In contrast to ranging, \oursystem exhibits a strong decoding ability. BER remains below $8\%$ even when the receiver is moved to a distance of $5$ m.

A considerable difference exists between ranging and decoding because decoding has a loose requirement for preamble alignment. In \oursystem, each bit has an interval of $5$ ms. Our study implies that a bit can still be decoded even when a misalignment of $2$ ms exists  (\ie involving $40\%$ of the samples)  because energy accumulation  across $60\%$ of the samples for a bit is sufficient for bit decision. By contrast, a misalignment of $2$ ms causes a $344.35\times0.002=68.87$ cm error in ranging. Background noise can easily obscure a few samples in the preamble. 

\subsection{Evaluation in Two-Dimension}

Then, we evaluate \oursystem in two dimensions with respect to its localization accuracy. 
 
\subsubsection{Accuracy in Localization}

We compare \oursystem's localization accuracy with that of TUPS, which uses ultrasonic sensor operating at ultrasonic bands as its receiver. We perform $150$ experimental trials. In each trial, the receiver is placed randomly in the evaluation environment. Fig.~\ref{fig:2d-comparison} plots the CDF of the 2D localization error. We observe the followings:

$\bullet$ \oursystem achieves a median accuracy of $4.59$ cm and a $90^{th}$ percentile accuracy of $14.57$ cm in 2D localization. By contrast, the median and $90^{th}$ percentile accuracy of the TUPS are $3.51$ cm and $15.13$ cm, respectively. These results are consistent with Dolphin's implementation~\cite{hazas2002novel,hazas2003high} which also reports a median localization accuracy of $3$ cm. This result demonstrates that \oursystem can achieve nearly the same accuracy as the traditional UPS even if it works at the second-order harmonics.   Interestingly,  the 2D localization accuracy of \oursystem behaves slightly better than that under 1D condition,  because trilateration in 2D uses the difference in ToA. Many uncertain common variables (\eg sync delay at \ubeacons or audio processing at the receiver) will be canceled out from the difference.

$\bullet$ \oursystem and Dolphin achieve more than $10\times$ improvement for the mean accuracy over Cricket~\cite{priyantha2000cricket,miu2002design}, which reports  a mean accuracy of $30$ cm. This is due to the rapid development of hardware, which provides considerably higher resolution in sampling frequency compared with that in a decade ago.

In summary,  adopting the second-order harmonic as localization media can achieve the same accuracy as that with the first-order. \oursystem even exceeds some past UPSs.

\subsubsection{Impacts of Parameters}

Next, we evaluate the localization accuracy as a function of different system parameters: 

\textbf{Impact of Bandwidth.} We evaluate \oursystem's localization accuracy as a function of bandwidth. The \cbeacon sweeps the spectrum with a constant slope.  Thus, increasing the length of preambles of the \ubeacons' beacons is equivalent to increasing the bandwidth of the downconverted beacons. Fig.~\ref{fig:2d-bandwidth} shows the impact of bandwidth on localization accuracy. The plot demonstrates that the accuracy monotonically improves with increased bandwidth. In particular, if \oursystem uses a $2$ kHz bandwidth, then the median error reaches $13.13$ cm and rapidly drops below $10$ cm for bandwidths larger than $4$ kHz.  This result is attributed to increased bandwidth, which provides finer granularity in separating the LOS path from echoes. In our default setting, $4$ kHz is adopted to balance time delay and accuracy.

\textbf{Impact of Receiver Diversity.} The nonlinearity effect of a microphone system is a key in \oursystem for ultrasonic downconversion. A natural question is whether this effect can be applied to different model receivers. To this end, we test 2D accuracy across four typical smart-phones.  The nonlinearity effect of microphone system is a key in \oursystem for the ultrasonic downconversion. A natural question is whether this effect could work across different model receivers? To this end, we test the 2D accuracy across five mainstream smart phones. 
A slight difference exists among these devices. The errors of the iPhone series's are worse than those of the others because their microphone systems are better in suppressing in nonlinearity. Even so, we observe that the nonlinearity effect still exists as a ubiquitous phenomenon.
 
\textbf{Impact of Environment.} We repeat experiments under different environments, such as playing music, noisy lecture room and narrow space, where the space is full of echoes at different frequencies. The results are shown in the top of Fig.~\ref{fig:bench-environment}.  We could see that there are about $5$ cm changes in the median accuracy. The worst case happens when playing music where the ambient sound might contain some high-frequency items. However, \oursystem has a strong ability to fight off the echoes in narrow space due to the DCSS technique. 

\subsubsection{Energy Consumption}

Finally, we present the energy consumption of a \ubeacon across with respect to components and cycles respectively. Fig.~\ref{fig:bench-energy-comp} shows the accumulated  energy of the two main components,  which are persistently consumed during one second. It can be seen that the WiFi communication for the PTP sync consumes $4\times$ energy than the transducer. The results show that wireless transceiver is indeed an energy hog in smart devices, which is consistent with many previous reports. Next, we estimate the actual energy consumption during one second regarding to the broadcast cycle (for localization) and sync cycle (for clock synchronization). The  results are shown in Fig.~\ref{fig:bench-energy-cycles}. It is clear that the low-duty cycle would save much more energy, especially when considering PTP sync. Specifically, a \ubeacon can save $58\%$ energy when adjusting its sync cycle from five seconds to one minute. With respect to our current settings (\ie $1$ s sync and $3$ s broadcast), each \ubeacon can work about $5$ months. If the PTP sync cycle is increased to $64$ s, the life can be prolonged to $8$ months.

\begin{figure}[!t]
  	\centering
  	\includegraphics[height = 2.5cm]{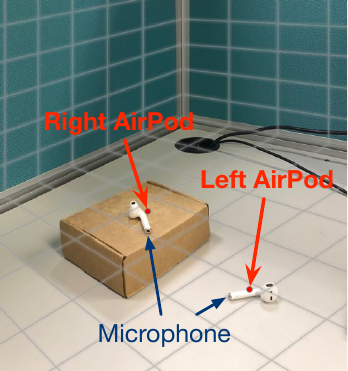}%
	\includegraphics[height = 2.5cm]{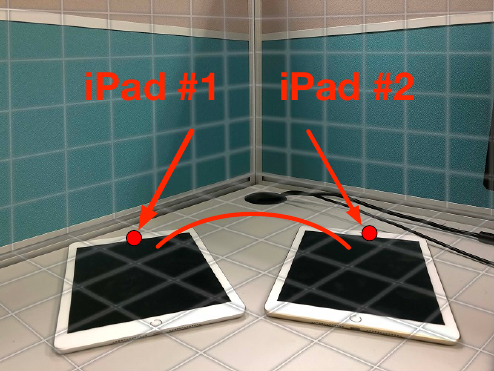}%
	\includegraphics[height = 2.5cm]{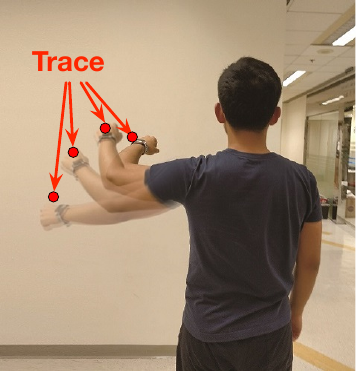}%
	\caption{\oursystem in real-world applications. \textnormal{(a) shows how \oursystem enables finding AirPods. (b) shows how \oursystem enables pairing iPads. (c) shows how \oursystem enables tracking hand arm through an Apple Watch.}}
   	\label{fig:vision}
\end{figure}

\subsection{Evaluation in Three-Dimension}

Finally, we vision three practical applications which can benefit from the high accuracy of \oursystem. They qualitatively test 3D localization accuracy in several real-world applications, as shown in Fig.~\ref{fig:vision}. 

\textbf{Finding AirPods}. This application shows to find a pair of AirPods (Ver. 1). Each AirPod integrates two microphones, but we are  allowed to access the bottom one only. Its sampling frequency is limited to $16$ kHz. For the better performance, we correspondingly adjust the chirp of \cbeacon to ensure the downconverted spectrum falls into $2\sim 7$ kHz. 

\textbf{Pairing iPads}. \oursystem could provide fundamental location service to VR-system. This application shows to pair two iPads once their distance is less than $10$ cm.

\textbf{Tracking Hand-Arm}. We apply \oursystem into tracking a human's hand-arm  through an Apple Watch (Ver. 3). The watch's highest frequency is up to $22$ kHz. However, we find that the built-in voice recorder uses M4A format, which filters the frequencies above $16$ kHz. This is also an important reason why we set the downconverted spectrum to $5\sim 15$ kHz, which can adapt to any kinds of formats.

\section{Limitations \& Conclusion}
\label{section:conclusion}

We present \oursystem, a system that enables previous UPSs to work in today's smart devices. We use the nonlinearity effect as an ultrasonic downconversion approach to pull down the frequencies of ultrasonic beacons. To do so, we invent two types of beacon devices, \ubeacon and \cbeacon. We believe that the implications of such a solution can pave the way for exciting new directions for the exploration of the acoustic community. 

However, the current prototype still has a couple of limitations. We outline them for future work.

\textbf{Relatively short range.} Although we have demonstrated that three-transducer made \ubeacon can achieve an effective range of $6$ m, the range is still far shorter than those of RF-based solutions. Actually, the short operating range is a common drawback of sound based localization system because acoustic wave attenuates faster than electromagnetic wave. However, the advantage of \oursystem is in its extremely high accuracy.  Our solution is more competitive for accuracy-sensitive applications, such as AR/VR systems, localizing tiny IoT devices, indoor navigation, and so on. Another reason that our solution is limited in the range is that we engineer the system using off-the-shelf ultrasonic transducers for saving cost. We can explore ultrasonic speakers like ALPS~\cite{lazik2015alps} to design long-ranged \ubeacons if necessary. 

\textbf{Pre-deployment.}   \oursystem requires pre-deploying beacon devices. Actually, all indoor localization solutions must use anchor devices like \ubeacons to establish the localization coordinate system, which is an unavoidable step.

\textbf{Post-maintenance.} Since all \ubeacon devices are battery-driven in \oursystem,  users are required to update batteries every a few months. Two methods can be utilized to mitigate this issue. First,  low-duty cycle algorithm is able to significantly reduce the energy consumption as shown in the evaluation. For example, all \ubeacons transit to low-powered silent states and are waken up by the smart devices when used; second, the recent advances in the wirelessly charging also provide a direction to design battery-free \ubeacons in our future work.

\section*{Acknowledgments}

The research is  supported by NSFC  General Program (NO. 61572282),   UGC/ECS (NO. 25222917), Shenzhen Basic Research Schema (NO. JCYJ20170818104855702), and Alibaba Innovative Research Program. 
We thank all the anonymous reviewers and the shepherd, Dr. Rajalakshmi Nandakumar, for their valuable comments and helpful suggestions.

\let\oldbibliography\thebibliography
\renewcommand{\thebibliography}[1]{%
  \oldbibliography{#1}%
  \setlength{\itemsep}{6pt}%
}

\newpage
{\small
\bibliographystyle{IEEEtran}
\bibliography{ups,tagoram,tagscreen,igps} 

\begin{thebibliography}{10}
\providecommand{\url}[1]{#1}
\csname url@samestyle\endcsname
\providecommand{\newblock}{\relax}
\providecommand{\bibinfo}[2]{#2}
\providecommand{\BIBentrySTDinterwordspacing}{\spaceskip=0pt\relax}
\providecommand{\BIBentryALTinterwordstretchfactor}{4}
\providecommand{\BIBentryALTinterwordspacing}{\spaceskip=\fontdimen2\font plus
\BIBentryALTinterwordstretchfactor\fontdimen3\font minus
  \fontdimen4\font\relax}
\providecommand{\BIBforeignlanguage}[2]{{%
\expandafter\ifx\csname l@#1\endcsname\relax
\typeout{** WARNING: IEEEtran.bst: No hyphenation pattern has been}%
\typeout{** loaded for the language `#1'. Using the pattern for}%
\typeout{** the default language instead.}%
\else
\language=\csname l@#1\endcsname
\fi
#2}}
\providecommand{\BIBdecl}{\relax}
\BIBdecl

\bibitem{shu2015last}
Y.~Shu, K.~G. Shin, T.~He, and J.~Chen, ``Last-mile navigation using
  smartphones,'' in \emph{Proc. of ACM MobiCom}, 2015.

\bibitem{haverinen2009global}
J.~Haverinen and A.~Kemppainen, ``Global indoor self-localization based on the
  ambient magnetic field,'' \emph{Robotics and Autonomous Systems}, vol.~57,
  no.~10, pp. 1028--1035, 2009.

\bibitem{zhang2017pulsar}
C.~Zhang and X.~Zhang, ``Pulsar: Towards ubiquitous visible light
  localization,'' in \emph{Proc. of ACM MobiCom}, 2017.

\bibitem{zhu2017enabling}
S.~Zhu and X.~Zhang, ``Enabling high-precision visible light localization in
  today's buildings,'' in \emph{Proc. of ACM MobiSys}, 2017.

\bibitem{wang2007rfid}
C.~Wang, H.~Wu, and N.-F. Tzeng, ``Rfid-based 3-d positioning schemes,'' in
  \emph{Proc. of IEEE INFOCOM}, 2007.

\bibitem{wang2013dude}
J.~Wang and D.~Katabi, ``Dude, where's my card?: Rfid positioning that works
  with multipath and non-line of sight,'' in \emph{Proc. of ACM SIGCOMM}, 2013.

\bibitem{wang2013rf}
J.~Wang, F.~Adib, R.~Knepper, D.~Katabi, and D.~Rus, ``Rf-compass: robot object
  manipulation using rfids,'' in \emph{Proc. of ACM MobiCom}, 2013.

\bibitem{ma20163d}
Y.~Ma, X.~Hui, and E.~C. Kan, ``3d real-time indoor localization via broadband
  nonlinear backscatter in passive devices with centimeter precision,'' in
  \emph{Proc. of ACM MobiCom}, 2016.

\bibitem{ma2017minding}
Y.~Ma, N.~Selby, and F.~Adib, ``Minding the billions: Ultra-wideband
  localization for deployed rfid tags,'' in \emph{Proc. of ACM MobiCom}, 2017.

\bibitem{xiong2013arraytrack}
J.~Xiong and K.~Jamieson, ``Arraytrack: A fine-grained indoor location
  system.'' in \emph{Proc. of USENIX NSDI}, 2013.

\bibitem{chen2012fm}
Y.~Chen, D.~Lymberopoulos, J.~Liu, and B.~Priyantha, ``Fm-based indoor
  localization,'' in \emph{Proc. of ACM MobiSys}, 2012.

\bibitem{bahl2000radar}
P.~Bahl and V.~N. Padmanabhan, ``Radar: An in-building rf-based user location
  and tracking system,'' in \emph{Proc. of IEEE INFOCOM}, 2000.

\bibitem{chintalapudi2010indoor}
K.~Chintalapudi, A.~Padmanabha~Iyer, and V.~N. Padmanabhan, ``Indoor
  localization without the pain,'' in \emph{Proc. of ACM MobiCom}, 2010.

\bibitem{johnson2012localization}
T.~A. Johnson and P.~Seeling, ``Localization using bluetooth device names,'' in
  \emph{Proc. of ACM MobiHoc}, 2012.

\bibitem{azizyan2009surroundsense}
M.~Azizyan, I.~Constandache, and R.~Roy~Choudhury, ``Surroundsense: mobile
  phone localization via ambience fingerprinting,'' in \emph{Proc. of ACM
  MobiCom}, 2009.

\bibitem{tan2013sound}
W.-T. Tan, M.~Baker, B.~Lee, and R.~Samadani, ``The sound of silence,'' in
  \emph{Proc. of ACM Sensys}, 2013.

\bibitem{gao2013zifind}
Y.~Gao, J.~Niu, R.~Zhou, and G.~Xing, ``Zifind: Exploiting cross-technology
  interference signatures for energy-efficient indoor localization,'' in
  \emph{Proc. of IEEE INFOCOM}, 2013.

\bibitem{adib20143d}
F.~Adib, Z.~Kabelac, D.~Katabi, and R.~C. Miller, ``3d tracking via body radio
  reflections.'' in \emph{NSDI}, vol.~14, 2014, pp. 317--329.

\bibitem{oxygen}
``{MIT Project Oxygen},'' \url{http://oxygen.csail.mit.edu}.

\bibitem{priyantha2000cricket}
N.~B. Priyantha, A.~Chakraborty, and H.~Balakrishnan, ``The cricket
  location-support system,'' in \emph{Proc. of ACM MobiCom}, 2000.

\bibitem{want1992active}
R.~Want, A.~Hopper, V.~Falcao, and J.~Gibbons, ``The active badge location
  system,'' \emph{ACM Transactions on Information Systems}, vol.~10, no.~1, pp.
  91--102, 1992.

\bibitem{priyantha2001cricket}
N.~B. Priyantha, A.~K. Miu, H.~Balakrishnan, and S.~Teller, ``The cricket
  compass for context-aware mobile applications,'' in \emph{Proc. of ACM
  MobiCom}, 2001.

\bibitem{smith2004tracking}
A.~Smith, H.~Balakrishnan, M.~Goraczko, and N.~Priyantha, ``Tracking moving
  devices with the cricket location system,'' in \emph{Proc. of ACM MobiSys},
  2004.

\bibitem{lazik2012indoor}
P.~Lazik and A.~Rowe, ``Indoor pseudo-ranging of mobile devices using
  ultrasonic chirps,'' in \emph{Proc. of ACM Sensys}, 2012.

\bibitem{lazik2015ultrasonic}
P.~Lazik, N.~Rajagopal, B.~Sinopoli, and A.~Rowe, ``Ultrasonic time
  synchronization and ranging on smartphones,'' in \emph{Proc. of IEEE RTAS},
  2015.

\bibitem{borriello2005walrus}
G.~Borriello, A.~Liu, T.~Offer, C.~Palistrant, and R.~Sharp, ``Walrus: wireless
  acoustic location with room-level resolution using ultrasound,'' in
  \emph{Proc. of ACM MobiSys}, 2005.

\bibitem{tarzia2011indoor}
S.~P. Tarzia, P.~A. Dinda, R.~P. Dick, and G.~Memik, ``Indoor localization
  without infrastructure using the acoustic background spectrum,'' in
  \emph{Proc. of ACM MobiSys}, 2011.

\bibitem{peng2007beepbeep}
C.~Peng, G.~Shen, Y.~Zhang, Y.~Li, and K.~Tan, ``Beepbeep: a high accuracy
  acoustic ranging system using cots mobile devices,'' in \emph{Proc. of ACM
  Sensys}, 2007.

\bibitem{hazas2002novel}
M.~Hazas and A.~Ward, ``A novel broadband ultrasonic location system,'' in
  \emph{Proc. of ACM Ubicomp}, 2002.

\bibitem{mccarthy2006accessible}
M.~McCarthy, P.~Duff, H.~L. Muller, and C.~Randell, ``Accessible ultrasonic
  positioning,'' \emph{IEEE Pervasive Computing}, vol.~5, no.~4, pp. 86--93,
  2006.

\bibitem{yang2011detecting}
J.~Yang, S.~Sidhom, G.~Chandrasekaran, T.~Vu, H.~Liu, N.~Cecan, Y.~Chen,
  M.~Gruteser, and R.~P. Martin, ``Detecting driver phone use leveraging car
  speakers,'' in \emph{Proc. of ACM MobiCom}, 2011.

\bibitem{mandal2005beep}
A.~Mandal, C.~V. Lopes, T.~Givargis, A.~Haghighat, R.~Jurdak, and P.~Baldi,
  ``Beep: 3d indoor positioning using audible sound,'' in \emph{Proc. of IEEE
  CCNC}, 2005.

\bibitem{zhang2017dolphinattack}
G.~Zhang, C.~Yan, X.~Ji, T.~Zhang, T.~Zhang, and W.~Xu, ``Dolphinattack:
  Inaudible voice commands,'' in \emph{Proc. of ACM SIGSAC}, 2017.

\bibitem{roy2017backdoor}
N.~Roy, H.~Hassanieh, and R.~Roy~Choudhury, ``Backdoor: Making microphones hear
  inaudible sounds,'' in \emph{Proc. of ACM MobiSys}, 2017.

\bibitem{roy2018inaudible}
N.~Roy, S.~Shen, H.~Hassanieh, and R.~R. Choudhury, ``Inaudible voice commands:
  The long-range attack and defense,'' in \emph{Proc. of USENIX NSDI}, 2018.

\bibitem{lee2015chirp}
H.~Lee, T.~H. Kim, J.~W. Choi, and S.~Choi, ``Chirp signal-based aerial
  acoustic communication for smart devices,'' in \emph{Proc. of IEEE INFOCOM},
  2015.

\bibitem{lin2017tagscreen}
Q.~Lin, L.~Yang, and Y.~Liu, ``Tagscreen: Synchronizing social televisions
  through hidden sound markers,'' in \emph{Proc. of IEEE INFOCOM}, 2017.

\bibitem{harter2002anatomy}
A.~Harter, A.~Hopper, P.~Steggles, A.~Ward, and P.~Webster, ``The anatomy of a
  context-aware application,'' \emph{Wireless Networks}, vol.~8, no. 2/3, pp.
  187--197, 2002.

\bibitem{hazas2003high}
M.~Hazas and A.~Ward, ``A high performance privacy-oriented location system,''
  in \emph{Proc. of IEEE PerCom}, 2003.

\bibitem{filonenko2010investigating}
V.~Filonenko, C.~Cullen, and J.~Carswell, ``Investigating ultrasonic
  positioning on mobile phones,'' in \emph{Proc. of IEEE IPIN}, 2010.

\bibitem{nandakumar2015contactless}
R.~Nandakumar, S.~Gollakota, and N.~Watson, ``Contactless sleep apnea detection
  on smartphones,'' in \emph{Proc. of ACM MobiSys}, 2015.

\bibitem{lazik2015alps}
P.~Lazik, N.~Rajagopal, O.~Shih, B.~Sinopoli, and A.~Rowe, ``Alps: A bluetooth
  and ultrasound platform for mapping and localization,'' in \emph{Proc. of ACM
  Sensys}, 2015.

\bibitem{rossi2013roomsense}
M.~Rossi, J.~Seiter, O.~Amft, S.~Buchmeier, and G.~Tr{\"o}ster, ``Roomsense: an
  indoor positioning system for smartphones using active sound probing,'' in
  \emph{Proc. of ACM AH}, 2013.

\bibitem{lorincz2005motetrack}
K.~Lorincz and M.~Welsh, ``Motetrack: A robust, decentralized approach to
  rf-based location tracking,'' in \emph{International Symposium on
  Location-and Context-Awareness}.\hskip 1em plus 0.5em minus 0.4em\relax
  Springer, 2005, pp. 63--82.

\bibitem{lu2009soundsense}
H.~Lu, W.~Pan, N.~D. Lane, T.~Choudhury, and A.~T. Campbell, ``Soundsense:
  scalable sound sensing for people-centric applications on mobile phones,'' in
  \emph{Proc. of ACM MobiSys}, 2009.

\bibitem{tung2015echotag}
Y.-C. Tung and K.~G. Shin, ``Echotag: Accurate infrastructure-free indoor
  location tagging with smartphones,'' in \emph{Proc. of ACM MobiCom}, 2015.

\bibitem{panchpor2018survey}
A.~A. Panchpor, S.~Shue, and J.~M. Conrad, ``A survey of methods for mobile
  robot localization and mapping in dynamic indoor environments,'' in
  \emph{Proc. of IEEE SPACES}, 2018.

\bibitem{jeub2012noise}
M.~Jeub, C.~Herglotz, C.~Nelke, C.~Beaugeant, and P.~Vary, ``Noise reduction
  for dual-microphone mobile phones exploiting power level differences,'' in
  \emph{Proc. of IEEE ICASSP}, 2012.

\bibitem{chen2017background}
Y.-Y. Chen and J.-H. Zhang, ``Background noise reduction design for dual
  microphone cellular phones: Robust approach,'' \emph{IEEE/ACM Transactions on
  Audio, Speech, and Language Processing}, vol.~25, no.~4, pp. 852--862, 2017.

\bibitem{martin2001noise}
R.~Martin, ``Noise power spectral density estimation based on optimal smoothing
  and minimum statistics,'' \emph{IEEE Transactions on speech and audio
  processing}, vol.~9, no.~5, pp. 504--512, 2001.

\bibitem{jeub2011robust}
M.~Jeub, C.~Nelke, H.~Kr{\"u}ger, C.~Beaugeant, and P.~Vary, ``Robust
  dual-channel noise power spectral density estimation,'' in \emph{Proc. of
  IEEE EURASIP}, 2011.

\bibitem{horowitz1989art}
P.~Horowitz and W.~Hill, \emph{The art of electronics}.\hskip 1em plus 0.5em
  minus 0.4em\relax Cambridge Univ. Press, 1989.

\bibitem{TI-PTP}
``{TI PTP Application Report},''
  \url{http://www.ti.com/lit/an/snla098a/snla098a.pdf }.

\bibitem{brown1992introduction}
R.~G. Brown, P.~Y. Hwang \emph{et~al.}, \emph{Introduction to random signals
  and applied Kalman filtering}.\hskip 1em plus 0.5em minus 0.4em\relax Wiley
  New York, 1992, vol.~3.

\bibitem{yang2014tagoram}
L.~Yang, Y.~Chen, X.-Y. Li, C.~Xiao, M.~Li, and Y.~Liu, ``Tagoram: real-time
  tracking of mobile rfid tags to high precision using cots devices,'' in
  \emph{Proc. of ACM MobiCom}, 2014.

\bibitem{xiong2012arraytrack}
J.~Xiong and K.~Jamieson, ``Arraytrack: a fine-grained indoor location
  system,'' in \emph{Proc. of USENIX NSDI}, 2012.

\bibitem{estimote}
``{Bluetooth iBeacon},'' \url{https://estimote.com/}.

\bibitem{transducer-datasheet}
``{Transducer Datasheet},''
  \url{https://www.murata.com/products/productdetail?partno=MA40S4S}.

\bibitem{esp32}
``{ESP32-DevKitC},''
  \url{https://www.espressif.com/en/products/hardware/esp32-devkitc/overview}.

\bibitem{vifa}
``{Ultrasonic Dynamic Speaker Vifa},'' \url{http://www.avisoft.
  com/usg/vifa.htm}.

\bibitem{miu2002design}
A.~K.~L. Miu, ``Design and implementation of an indoor mobile navigation
  system,'' Ph.D. dissertation, Massachusetts Institute of Technology, 2002.

\end{thebibliography}
}

\end{document}